\documentclass[12pt]{iopart}
\usepackage{graphicx}
\begin{document}



\title{Probing the Nuclear Symmetry Energy with Heavy Ion Collisions}

\author{M Di Toro$^1$ $^2$, V Baran 
$^3$, M Colonna$^1$, V Greco $^1$ $^2$ }

\address{$^1$ Laboratori Nazionali del Sud INFN, I-95123 Catania, Italy} 
\address{$^2$ Physics and Astronomy Dept., University of Catania}
\address{$^3$ Physics Faculty, Univ. of Bucharest and NIPNE-HH, Romania}

\ead{ditoro@lns.infn.it}

\begin{abstract}

Heavy Ion Collisions ($HIC$) represent a unique tool to probe the 
in-medium
nuclear interaction in regions away from saturation. In this report we 
present a 
selection of new reaction observables in dissipative collisions 
particularly sensitive to the
symmetry term of the nuclear Equation of State ($Iso-EoS$).
We will first discuss the 
Isospin Equilibration Dynamics. At low energies this
 manifests via
the recently observed Dynamical Dipole Radiation, due
to a collective neutron-proton oscillation with the symmetry 
term acting 
as a restoring force. At higher beam energies Iso-EoS
effects will be seen in an Isospin Diffusion mechanism, via 
Imbalance Ratio Measurements, in particular from  
correlations to the total kinetic energy loss. 
For fragmentation reactions in central events we suggest to 
look at the coupling between isospin distillation and radial flow. In Neck 
Fragmentation reactions 
important $Iso-EoS$ information can be obtained 
from fragment isospin content, velocity and alignement correlations.

The high density symmetry term can be probed from
 isospin effects on 
heavy ion reactions
at relativistic energies (few $AGeV$ range),
 in particular for high transverse momentum selections 
of the reaction products.
Rather isospin sensitive 
observables are proposed from nucleon/cluster 
emissions, collective flows
and meson 
production. The possibility  to shed light 
on the controversial neutron/proton effective 
mass splitting in asymmetric 
matter is also suggested. 

A large symmetry repulsion at high 
baryon density
will also lead to an ``earlier'' hadron-deconfinement transition
in n-rich matter. 
The binodal transition line of the 
($T,\rho_B$) diagram is lowered to a region accessible through heavy ion 
collisions in the energy range of the new planned facilities, e.g. the
$FAIR/NICA$ projects. Some observable 
effects of the formation of a Mixed Phase are suggested, in particular a 
{\it Neutron Trapping} mechanism.
The dependence of the results on a suitable treatment of 
the isovector part of the interaction in effective QCD Lagrangian approaches
is critically discussed.

We stress the interest of this study in nuclear astrophysics, in particular 
for supernovae explosions and neutron star structure, where the 
knowledge of the
$Iso-EoS$ is important at low as well as at high baryon density.
\end{abstract}
\submitto{\JPG}
\maketitle

\section{Introduction: The Elusive Symmetry Term of the EoS}

The study of the behavior of nuclear matter under several conditions of 
density and temperature
is of crucial importance for the understanding of a large variety of 
phenomena,   
ranging from the structure of nuclei and their decay modes, up to the 
life and the
properties of massive stars. 
Mechanisms involving an enormous range of scales in size, characteristic 
time and energy, 
but all based on nuclear processes at fundamental level, are actually 
linked by the concept 
of the nuclear Equation of State ($EoS$) and the associated energy density 
functional.
In particular, the understanding of the properties of exotic nuclei, 
as well as
neutron stars and supernova dynamics, entails constraining the behavior 
of the nuclear symmetry 
energy, which describes the difference between the binding energy of 
symmetric matter (with
equal proton and neutron numbers, N=Z), and that of pure neutron matter.

Transient states of nuclear matter far from normal conditions can be created
in terrestrial laboratories via Heavy Ion Collisions ($HIC$). 
Many experimental and theoretical efforts 
have been devoted,
over the past 30 years, to the study of nuclear reactions from low to 
intermediate energies, as a 
possible tool to learn about the behavior of nuclear matter and its $EoS$. 
Relevant conclusions have been reached concerning the $EoS$ of symmetric 
matter
for densities up to five time the saturation value \cite{daniel02}.
More recently, the availability of neutron-rich and exotic beams has 
opened
the way to investigate, in laboratory conditions, new aspects of nuclear 
structure
and dynamics up to extreme ratios of neutron to proton numbers.
Thus it has become possible to explore the behavior of nuclear matter 
along a new
degree of freedom, the asymmetry $I = (N-Z)/(N+Z)$ (in the rest of the
review also defined as $\beta$ or $\alpha$), aiming at constraining 
the density
and/or temperature dependence of the symmetry energy  ($Iso-EoS$). 
Here we will review the Isospin Dynamics in $HIC$ from the Coulomb Barrier 
to Relativistic Energies.

 The symmetry energy $E_{sym}$ appears in the energy density
$\epsilon(\rho,\rho_3)\equiv \epsilon(\rho)+\rho E_{sym} (\rho_3/\rho)^2
 + O(\rho_3/\rho)^4 +..$, expressed in 
terms of total ($\rho=\rho_p+\rho_n$)
 and isospin ($\rho_3=\rho_p-\rho_n$) densities \cite{baranPR}. The 
symmetry term gets a
kinetic contribution directly from basic Pauli correlations and a potential
part from the highly controversial isospin dependence of the effective 
interactions. Both at sub-saturation and supra-saturation
densities, predictions based of the 
existing many-body techniques diverge 
rather widely, see \cite{fuchswci,fantoni08}. 
We  recall  that the knowledge of the 
 $EoS$  of asymmetric matter is very important at low 
densities, in nuclear structure ( neutron skins,
 pigmy resonances, refs. \cite{gsi07,pieka06,trippa08,carbone10}, in reactions 
(neutron distillation 
in fragmentation \cite{baran98}, charge equilibration \cite{tsang04}),
 in astrophysics (neutron star formation and crust, \cite{horopieka02,
steiner05}) as well as at high 
densities in relativistic heavy ion reactions (isospin flows \cite{greco03}, 
particle
production \cite{baranPR,baoPR08,ferini05,ferini06}), in compact star  
(neutron star mass-radius relation, cooling, hybrid structure, formation
of black holes, \cite{page06,latpra07,prakash07,baldo07}) and for
fundamental properties of strong interacting systems (transition to
new phases of the matter, \cite{muller,ditorodec}).

Several observables which
are sensitive to 
the $Iso-EoS$ and testable
experimentally, have been suggested
\cite{baranPR,baoPR08,colonnaPRC57,baoIJMPE7,Isospin01,baran2004,
wcineck,WCI_betty,sheyen10}.
We take advantage of new opportunities in 
theory (development of rather reliable 
microscopic transport codes for $HIC$)
 and in experiments (availability of very asymmetric 
radioactive beams, 
improved possibility of measuring event-by-event correlations) to 
present
new results that are constraining the existing effective interaction 
models. We will 
discuss dissipative collisions in a wide range of beam 
energies, 
 from just above the 
Coulomb barrier up to the $AGeV$ range. Isospin effects
on the chiral/deconfinement 
transition at high baryon density will be also
analyzed.
Low to Fermi energies
 will bring 
information on the symmetry term around (below) normal density, 
while intermediate energies 
will probe high density regions.
The transport codes are based on 
mean field theories, with 
correlations included via hard nucleon-nucleon
elastic and inelastic collisions and via 
stochastic forces, selfconsistently
evaluated from the mean phase-space trajectory, see 
\cite{baranPR,guarneraPLB373,colonnaNPA642,chomazPR}. 
Stochasticity is 
essential in 
order 
to get distributions as well as to allow the growth of dynamical 
instabilities. 

Relativistic collisions are described via a fully covariant transport 
approach, related to 
an effective field exchange model, where the relevant 
degrees of freedom of the nuclear 
dynamics are accounted for
 \cite{baranPR,ferini05,ferini06,liubo02,theo04,santini05}.
We will have a propagation of particles suitably dressed by self-energies
that will influence 
collective flows and in medium nucleon-nucleon inelastic 
cross sections. The construction of 
an $Hadron-EoS$ at high baryon and 
isospin densities will finally allow the possibility of 
developing a model of a 
hadron-deconfinement transition at high density for an asymmetric 
matter
\cite{ditorodec}. The problem of a correct treatment of the isospin in an
effective partonic $EoS$ will be stressed.

\section{The Stochastic Mean Field transport approach}
At low to intermediate energies the evolution of systems governed  
by a complex phase space
can be described
via  a transport equation (of the Boltzmann-Nordheim-Vlasov type) with a 
fluctuating term, 
the so-called
Boltzmann-Langevin equation (BLE) \cite{baranPR}
\begin{equation}
{{df}\over{dt}} = {{\partial f}\over{\partial t}} + \{f,H\} = I_{coll}[f] 
+ \delta I[f],
\end{equation}
 where $f({\bf r},{\bf p},t)$ is the one-body distribution function, 
or Wigner transform of the one-body density, 
$H({\bf r},{\bf p},t)$ the mean field Hamiltonian, 
$I_{coll}$ the two-body collision term (that accounts for the residual 
interaction) 
incorporating the Fermi statistics of the particles,
and 
$\delta I[f]$ the fluctuating part of the
collision integral \cite{Ayik,Randrup}.
The system is described in terms of the one-body distribution function $f$, 
however this function
may experience a stochastic evolution in response to the action of the 
fluctuating term $\delta I[f]$.
This is an effective way to insert again into the dynamics the effects of 
unknown correlations and of 
the loss of information caused by the projection of the many-body dynamics 
onto a much reduced subspace.
Here we will follow the approximate treatment to the BLE 
presented in Refs.\cite{guarneraPLB373,colonnaNPA642}, 
the Stochastic Mean Field (SMF) model, 
that consists in the implementation of stochastic spatial density 
fluctuations. 
The nuclear EoS, directly linked to the mean-field Hamiltonian $H$, can 
be written as: 
\begin{equation}
\frac{E}{A} (\rho,I) = \frac{E}{A}(\rho) +  E_{sym}(\rho)I^2 + O(I^4)
\end{equation}
We adopt a soft isoscalar EOS, $E/A(\rho)$, with compressibility modulus 
$K = 220~MeV$, 
which is favored from monopole oscillations in stable nuclei as well as
from flow studies \cite{daniel02}.

We will always test the sensitivity of our simulation results to
different choices of the density and momentum dependence of the
Isovector part of the Equation of State ($Iso-EoS$). In the non-relativistic
frame
the potential part of the symmetry energy, $C_{sym}(\rho)$, \cite{baranPR}:
\begin{equation}
E_{sym} = E_{sym}(kin)+E_{sym}(pot)\equiv 
\frac{\epsilon_F}{3} + C_{sym}(\rho)
\label{esym}
\end{equation}
 is tested by employing three different density
parametrizations of the symmetry potentials 
\cite{baranPR,colonnaPRC57,bar02},
one with a rapidly increasing behaviour 
with density, roughly proportional to $\rho^2$ ($Asy-superstiff$),
one with a linear increase of the potential  part
of the symmetry energy with density ($Asy-stiff$)
and one with a kind of saturation around normal
density ($Asy-soft$), even decreasing at higher densities.

\begin{figure}
\unitlength1cm
\begin{center}
\includegraphics[width=6.0cm]{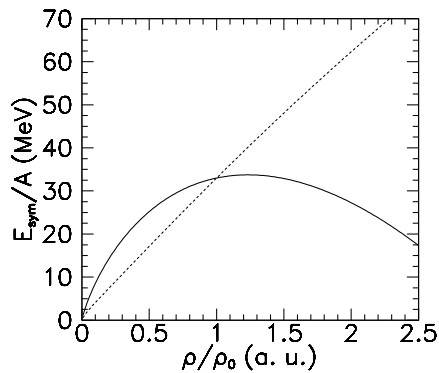}
\caption{Density dependence of the  symmetry energies  used in the simulations
presented here:  Asy-soft (solid) and Asy-stiff (dashed).
}
\end{center}
\label{esymdens}
\end{figure}

In particular, for
the symmetry term with the stronger density dependence,
\begin{equation}
E_{sym}\left( \rho \right) =a\cdot \left( \frac{\rho }{\rho _{0} } 
\right) ^{2/3}
+b\cdot \frac{2\left( \rho /\rho _{0} \right) ^{2} }{1+\left( \rho /\rho
_{0} \right)},
\label{eq3}
\end{equation}
where $\rho_0$ is the saturation density, a=12.7 MeV (fixed by the kinetic
contribution Eq.\ref{esym}) and b=19 MeV, to give a saturation value
of $31.7~MeV$.
The linear density dependence is simply given by:
\begin{equation}
E_{sym}\left( \rho \right) =a\cdot \left( \frac{\rho }{\rho _{0} } \right) ^
{2/3}
+ b \cdot (\rho / \rho_0),
\label{eq3new}
\end{equation}
For the symmetry term with weaker density dependence around saturation
\begin{equation}
E_{sym}\left( \rho \right) =a\cdot \left( \frac{\rho }{\rho _{0} } \right) ^
{2/3}
+ 240.9\rho  -819.1\rho ^{2},
\label{eq4}
\end{equation}
where a=12.7 MeV. 

The three parameterizations cross at normal density at the empirically known 
symmetry energy coefficient of the Bethe-Weizs\"acker formula.
As shown in Fig.1 the symmetry energy
at densities below the normal value 
is larger in the Asysoft case, 
while  above saturation  it is higher in the Asystiff cases.  
Hence in the low-density regime, that is the
region of interest for our analysis in heavy ion collisions at Fermi energies, 
isospin effects are expected to
be stronger in the Asysoft case.
Opposite expectations can be derived for relativistic collisions, where
high density regions will be probed in the early stage of the collision. 
In any case, for mechanisms sensitive to the density derivative of the 
symmetry energy an Asystiff-like behavior is more effective.

\subsection{Momentum Dependence of the Symmetry Potentials}

A  particular attention is devoted to 
the isospin effects on the momentum dependence of the symmetry potentials,
i.e. to observables sensitive to a different neutron/proton effective mass 
in asymmetric matter.
The problem of Momentum Dependence in the Isovector
channel ($Iso-MD$) is still very controversial and it would be extremely
important to get more definite experimental information,
see the refs. 
\cite{baranPR,baoPR08,rizzo04,ditoroAIP05}. 
Exotic beams at intermediate energies are
of interest in order to have high momentum particles and to test regions
of high baryon (isoscalar) and isospin (isovector) density during the
reaction dynamics. 

Our transport code has been implemented with 
a  generalized form of the effective interactions, which can be easily
reduced to Skyrme-like forces, with momentum dependent terms  also  in the
isovector channel  \cite{ditoroAIP05,rizzoj_th,rizzoPRC72,rizzo08},
i.e. with a different $(n,p)$ mean field momentum dependence.
The general structure of the  isoscalar and isovector Momentum Dependent (MD) 
 effective fields is
derived via an isospin asymmetric extension of the Gale-Bertsch-DasGupta  
(GBD)  force
 \cite{GBD,GalePRC41,BombaciNPA583,GrecoPRC59,BaoNPA735},  which  
corresponds to a Yukawian
 non-locality.
The isovector momentum dependence implies different effective masses
for protons and neutrons given as
$\frac{m^*_{\tau}}{m}=(1+\frac{m}{\hbar^2 p}\frac{\partial U_{\tau}}
{\partial p})^{-1}$, for $ p=p_{F,\tau}$, at fixed density. 
Thus our approach will allow to follow the dynamical
effect of opposite $n/p$ effective mass splitting while keeping the
same density dependence of the symmetry energy \cite{rizzo08,giordano10}.

\begin{figure}[htb]
\centering
\vskip 0.7cm
\includegraphics[width=8.0cm]{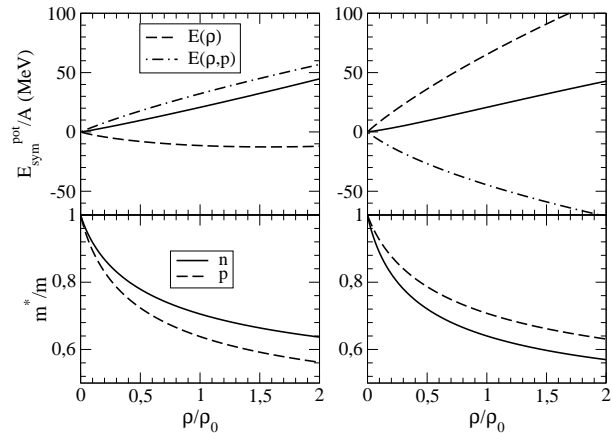}
\caption{Upper Panels: Density dependence of the potential symmetry energy 
(Solid Lines), in 
the $Asystiff$
choice. Dashed lines 
refer to
local contributions, dot-dash lines to momentum-dependent ones, see text. 
Left: 
 $m^*_n>m^*_p$ parametrization; Right: $m^*_n<m^*_p$ case.
Lower Panels: corresponding behavior of neutron/proton effective masses as a 
function 
of the density,
for an asymmetry $\beta=0.2$.}
\label{esymastar} 
\end{figure}

In fact when we use momentum-dependent interactions we have also contributions 
to the symmetry energy 
from the non-local terms. This is hown in Fig.\ref{esymastar} 
where we plot the density 
dependence of the potential part 
of the symmetry energy, 
in the Asy-stiff case,
for the two choices of the n/p mass splitting (solid lines, upper panels). 
We also separately report the 
contributions from 
the momentum-dependent, $E(\rho,p)$, and the density dependent, $E(\rho)$, 
part 
of the EoS, whose sum gives the total $E_{sym}^{pot}$ (the $C_{sym}(\rho)$
of Eq.(\ref{esym})). 
A change in the sign of the mass
splitting is related to opposite behaviors of these two contributions, 
exactly like it happens in
Skyrme-like forces, see sections (2.1-2.2) of ref.\cite{baranPR}. 
The lower panels show
the density dependence of the corresponding mass splitting, for an
asymmetry parameter $\beta=0.2$ (the $^{197}$Au asymmetry). In order to
probe the mass splitting effects on the heavy ion dynamics we have chosen
parametrizations that give almost opposite splittings at all densities.

\begin{figure}[h!]
\centering
\includegraphics[width=8.0cm,angle=0]{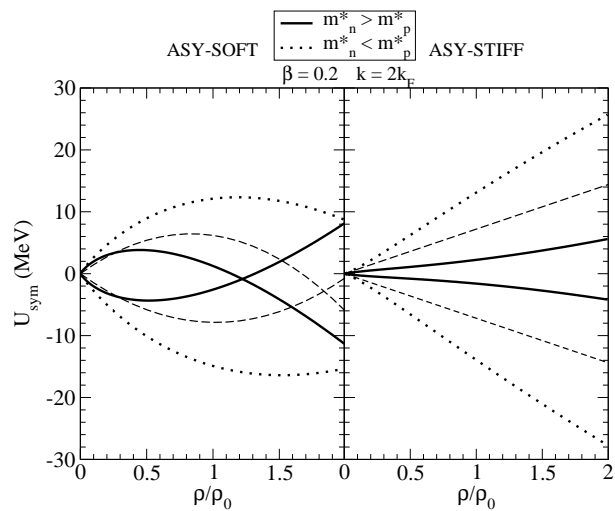}
\caption{Density dependence of neutron(upper)-proton(lower) potentials for
an asymmetry $\beta=0.2$ for the $Asysoft$(left) and $Asystiff$(right) 
choices. 
Dashed: No momentum dependence. Momentum dependent potentials at $k=2k_F$:
solid lines for the $m_{n}^{*} > m_{p}^{*}$ case, dotted lines for the opposite
$m_{n}^{*} < m_{p}^{*}$ choice.}
\label{Upot}
\end{figure}

In Fig.\ref{Upot} we present the density dependence of the neutron/proton 
symmetry potentials,
 for the two stiffness of the symmetry term, evaluated in the 
case without momentum 
dependence (dashed lines) and in the momentum dependence ($Iso-MD$) case 
for the  
$m_{n}^{*} < m_{p}^{*}$ 
(dotted) and the opposite $m_{n}^{*} > m_{p}^{*}$ (solid) choices.  We see 
that the momentum dependence
modifies the effect of the symmetry term stiffness on the nucleon 
potentials, with differences that
become more appreciable with increasing nucleon momenta. From this 
figure we can already
predict large effects of the effective mass splitting at high momenta.

\begin{figure}[h!]
\centering
\includegraphics[width=8.0cm,angle=0]{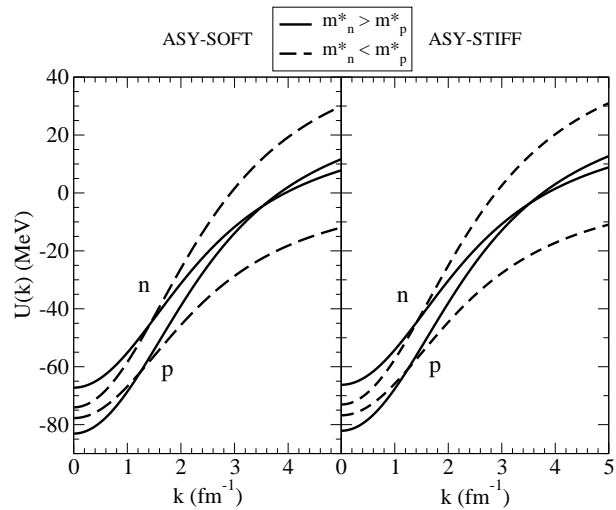}
\caption{Momentum dependence of neutron-proton potentials at saturation 
density and asymmetry $\beta=0.2$, for the two splitting choices 
$m_{n}^{*}<m_{p}^{*}$ (dashed) and 
$m_{n}^{*}>m_{p}^{*}$ (solid). Left panel: $Asysoft$ $Iso-Eos$. 
 Right panel: $Asystiff$ case}
\label{mdpot}
\end{figure}

This is shown more explicitly in Fig.\ref{mdpot} where we see the 
momentum dependence
of the neutron-proton potentials at saturation density for the two mass
splitting choices,
always for a "typical" $\beta=0.2$ asymmetry ($^{124}Sn, ^{197}Au$...).
The plot is for the $Asysoft$ (left panel) and the $Asystiff$ (right) 
symmetry term, and in fact it is not much different. Indeed
we can see also from the previous Fig.\ref{Upot}  
that at normal
density the difference between neutron and proton 
potentials is
almost the same for the two asy-stiffness, even in the case of 
$Iso-MD$ interactions..

The Figs.\ref{Upot} and \ref{mdpot} suggest the presence of 
interesting Isoscalar and 
Isovector $MD$ effects on the reaction dynamics: 
\begin{itemize}
\item 
{$Isoscalar$. In general the momentum dependence gives more 
attractive potentials at low 
momenta, $p<p_F$, and more repulsive at high $p$, $p>p_F$. 
In the reaction dynamics we expect
the more energetic nucleons to be fast emitted and to suffer 
less collisions \cite{GBD}. As a 
consequence we will 
have less stopping of the matter and less compression. 
The isoscalar EoS becomes stiffer
 when the momentum dependence is included.}
\item
{$Isovector$. Isospin effects on the momentum dependence imply 
different slopes around $p_F$
for neutrons and protons, as clearly shown in Fig.\ref{mdpot}, and so 
the larger repulsion
above $p_F$ is different. In the case $m_{n}^{*}<m_{p}^{*}$ the high 
momentum neutrons will
see a more repulsive field with respect to the high-$p$ protons. The 
opposite will happen
in the $m_{n}^{*}>m_{p}^{*}$ case. The fast nucleon emission will be 
directly affected: in the
$m_{n}^{*}<m_{p}^{*}$ case we expect a larger $n/p$ yield for nucleons 
emitted in central collisions
and a larger neutron $Squeeze-out$ (elliptic flow) in semicentral 
collisions in heavy ion reactions
at intermediate energies, in particular for high $p_t$ (transverse momentum) 
selections. 
In fact in the interacting, high density, early stage of the reaction dynamics
the pressure is built from
violent nucleon-nucleon collisions and the high $p_t$ particles will carry
the maximal information on high density and momentum dependence of the 
symmetry potentials. The azimuthal distributions (elliptic flows) will be
particularly affected since particles mostly retain their high transverse
momenta escaping along directions orthogonal to the reaction plane without
suffering much rescattering processes. 

In Sect.5 we will test those 
predictions
also for n-rich vs. n-poor light ions, like ($^3H$, $^3He$), easier to 
detect. Since, as already 
noted, the symmetry potentials are not very different in the 
Asystiff/Asysoft choice for the density 
range probed at intermediate energies, we can expect that 
the $Mass-Splitting$
effect could be even larger than the one related to the different 
stiffness of the symmetry term. 
}
\end{itemize}

\section{Isospin Equilibration in Low Energy Dissipative Collisions}

The presence of an Isovector Dipole Oscillation in the entrance channel 
dynamics has been suggested by several authors in order to account for 
the fast charge equilibration, and even for the fragment charge distributions,
in Deep Inelastic Collisions 
\cite{berlanger79,hernandez81,bonche81,ditgre85,suraud89,brink90}.
Since the oscillation is triggered by the mean field of a dilute Dinuclear 
Composite System this mechanism seems to appropriate for the study of
the symmetry term below saturation, which is acting as a restoring force.
A direct observation of the corresponding radiative emission would be 
then interesting. In fact the clear dynamical features of such 
{\it pre-equilibrium}
dipole mode (large deformations of the source, preferential oscillation on the
reaction plane) will allow to distinguish this radiation from the 
$\gamma$-emission of the statistical Giant Dipole Resonances ($GDR$) of
the final excited reaction products.

\subsection{The Prompt Dipole $\gamma$-Ray Emission}
The 
possibility of an entrance channel collective dipole, and corresponding
 radiation, 
due to an initial 
different N/Z distribution was predicted at the beginning
of the nineties 
\cite{ChomazNPA563,BortignonNPA583}. 
After several experimental evidences, in fusion as well 
as in deep-inelastic
reactions, \cite{PierrouPRC71,medea} and refs. therein,  
we have now a 
good understanding of the process and stimulating new perspectives
from the use of 
radioactive beams, to enhance the sensitivity to the
$Iso-EoS$.

\begin{figure}
\begin{center}
\includegraphics*[angle=-90,scale=0.33]{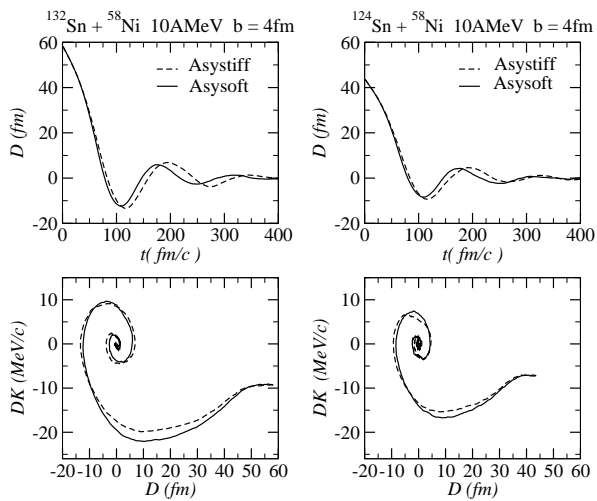}
\end{center}
\vskip -0.5cm
\caption{Dipole Dynamics at 10AMeV, $b=4fm$ centrality. 
Left 
Panels: Exotic ``132'' system. Upper: Time evolution of dipole moment 
D(t) in real 
space; 
Lower: Dipole phase-space correlation (see text).
Right Panels: same as before for the stable 
``124'' system.
Solid lines correspond to Asy-soft EoS, the dashed to Asy-stiff EoS.}
\label{dip}
\end{figure}

 In the first stages 
of dissipative reactions between colliding ions with
 different N/Z
ratios, a large amplitude 
dipole collective motion develops in the mean field of the composite
dinuclear system, the so-called Dynamical 
Dipole mode. This 
gives rise to a prompt $\gamma $-ray emission which 
depends:
 i) on the absolute
value of the intial dipole moment
\begin{eqnarray}
&&D(t= 0)= 
\frac{NZ}{A} \left|{R_{Z}}(t=0)- {R_{N}}(t=0)\right| =  \nonumber \\
&&\frac{R_{P}+R_{T}}{A}Z_{P}Z_{T}\left| (\frac{N}{Z})_{T}-(\frac{N}{Z})_{P}
\right|,
\label{indip}
\end{eqnarray}
being ${R_{Z}}= \frac {\Sigma_i x_i(p)}{Z}$ and
${R_{N}}=\frac 
{\Sigma_i x_i(n)}{N} $ the
center of mass of protons and of neutrons respectively, while 
R$_{P}$ and
R$_{T}$ are the
projectile and target radii; ii) on the fusion/deep-inelastic 
dynamics, which rules the formation of the dinuclear mean field;
 iii) on the symmetry term, below saturation, that is acting as a restoring
force.
A detailed description can consistently obtained in mean field transport 
approaches,
\cite{BrinkNPA372,BaranNPA600,SimenPRL86,BaranPRL87,simenel}.
We can follow the time evolution
of the dipole moment
in the 
$r$-space,
 $D(t)= \frac{NZ}{A} ({R_{Z}}- {R_{N}})$ and in
$p-$space, 
$DK(t)=(\frac{P_{p}}{Z}-\frac{P_{n}}{N})$, 
with $P_{p}$
($P_{n}$) center of mass in momentum 
space for protons (neutrons),
just the canonically conjugate momentum of the $D(t)$ 
coordinate,
i.e. as operators $[D(t),DK(t)]=i\hbar$ \cite{simenel}. 
A nice "spiral-correlation"
clearly 
denotes the collective nature
 of the mode, see Fig.\ref{dip}
We can directly
apply a 
bremsstrahlung approach,
 to the dipole evolution given from the Landau-Vlasov transport
\cite{BaranPRL87}, to estimate the ``prompt'' photon emission probability
($E_{\gamma}= \hbar \omega$):
\begin{equation}
\frac{dP}{dE_{\gamma}}= \frac{2 e^2}{3\pi \hbar c^3 E_{\gamma}}
 |D''(\omega)|^{2}  \label{brems},
\end{equation}
where $D''(\omega)$ is the Fourier transform 
of the dipole acceleration
$D''(t)$. We remark that in this way it is possible
to evaluate, 
in {\it absolute} values, the corresponding pre-equilibrium
photon emission.

In a  recent 
experiment the prompt dipole radiation has been 
investigated with
a $4 \pi$ gamma detector. 
A strong dipole-like photon angular distribution
$(\theta_\gamma)=W_0[1+a_2P_2(cos 
\theta_\gamma)]$, $\theta_\gamma$ being the 
angle between the emitted photon and  
the beam 
axis, has been observed,
 with the 
$a_2$ parameter close to $-1$, see \cite{medea}. 
At higher beam energies we expect a decrease of the direct dipole
radiation for two main reasons 
both due to the increasing importance of hard
NN collisions: i) a larger fast neutron 
emission that will equilibrate the
isospin during the dipole oscillation; ii) a larger 
damping of the
collective mode due to $np$ collisions.

The use of unstable neutron rich projectiles would largely increase the
effect, due 
to the possibility of larger entrance channel asymmetries
 \cite{ditoro_kaz07}.
In order to 
suggest proposals for the new $RIB$ facility $Spiral~2$, 
\cite{lewrio} we have studied 
fusion events in the reaction $^{132}Sn+^{58}Ni$ 
at $10AMeV$, \cite{ditoro_kaz07,spiral2}. 
We espect a $Monster$ 
Dynamical Dipole, the initial 
dipole moment $D(t=0)$ being of the 
order of 50fm, about two times the largest 
values probed so far, allowing a detailed study 
of the symmetry potential, 
below 
saturation,
responsible of the restoring force of the 
dipole oscillation and even 
of the damping,
 via the fast neutron emission. 

In Fig.	\ref{dip}
we report some global informations concerning the dipole mode
in entrance channel. 
In the Left-Upper panel we have the time evolution of 
the dipole moment $D(t)$
for the ``132'' system at $b=4fm$.
We notice the large amplitude of 
the first oscillation
but also the delayed dynamics for the Asy-stiff EoS related to a weaker 
isovector
restoring force.

The phase space correlation (spiraling) between $D(t)$
and $DK(t)$, is
reported in Fig.\ref{dip} 
(Left-Lower). 
It nicely points out a collective behavior which initiates very early,
with a 
dipole moment still close to the touching configuration value
reported above.
This can be 
explained by the fast formation of a well developed neck mean field
which sustains the 
collective
dipole oscillation in the dinuclear configuration.

The role of a large charge asymmetry between 
the two 
colliding nuclei can be seen from
Fig.\ref{dip} (Right Panels), where we show the analogous dipole 
phase space 
trajectories for the stable  
$^{124}Sn+^{58}Ni$ system at the same value of 
impact parameter and energy. 
A clear 
reduction of the 
collective behavior is 
evidenced. 

\begin{figure}
\begin{center}
\includegraphics*[scale=0.33]{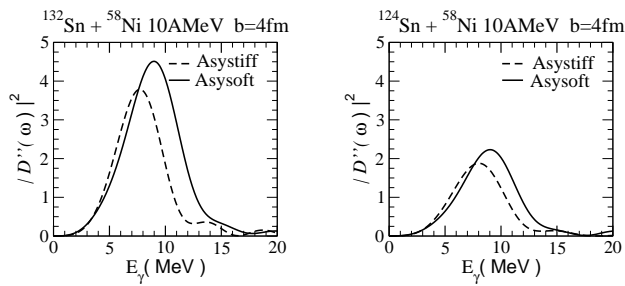}
\end{center}
\vskip -0.5cm
\caption{Left Panel, Exotic ``132'' system. Power spectra of 
the 
dipole acceleration at  $b=4$fm (in $c^2$ units).
Right Panel: Corresponding results for 
the stable ``124'' system.
Solid lines correspond to Asysoft EoS, the dashed to Asystiff 
EoS.}
\label{yield1}
\end{figure}

In Fig.\ref{yield1} (Left Panel) we report the power 
spectrum, 
$\mid D''(\omega) \mid^2$ in semicentral
``132'' reactions, for different 
$Iso-EoS$ choices.
The gamma multiplicity is simply related to it, see Eq.(\ref{brems}).
The corresponding results for the stable ``124''  system are drawn
in the Right Panel.
As expected from the larger initial charge asymmetry, we clearly see an 
increase 
of the Prompt 
Dipole Emission for the exotic
n-rich beam. Such entrance channel effect will be enhanced, 
allowing
a better observation of the Iso-EoS dependence.

We recall that
in the Asystiff case we have a weaker restoring force 
for the dynamical 
dipole
in the dilute ``neck'' region, where the symmetry energy is smaller
\cite{baranPR}.
This is reflected in
lower values of the centroids as well as in reduced total yields, as 
shown 
in Fig.\ref{yield1}. The sensitivity of $\omega_0$ to the stiffness 
of the symmetry
energy will 
be amplified by the increase of $D(t_0)$ when we use exotic,
more asymmetric beams. 

The 
prompt dipole radiation angular distribution is the result of the 
interplay between the 
collective oscillation life-time and the dinuclear 
rotation. In this sense we also expect  a 
sensitivity to the $Iso-EoS$ of the
anisotropy, in particular for high spin event selections. 
E.g. we remind that in the Asy-stiff case we have a delayed
onset of the collective dipole (see Fig.\ref{dip}), the emitting system will
get more rotation and the angular anisotropy will be reduced
\cite{dipang08}.

In the Asysoft 
choice we expect also larger widths of the
"resonance" due to the larger fast neutron 
emission.
We note the opposite effect of the Asy-stiffness on neutron vs proton 
emissions.
The latter point is important even for the possibility of an independent 
test just
measuring 
the $N/Z$ of the pre-equilibrium nucleon emission, \cite{Famiano,pfabe_iwm}.

In general the collective dipole mechanism for charge equilibration
will be important in a limited range of beam energies \cite{BaranPRL87}.
We must be well above the Coulomb Barrier in order to have a fast
formation of the dinuclear mean field, but when the direct nucleon-nucleon 
collisions will be more frequent we expect a rapid quenching of the isovector
collective mode. In Sect.4 we will discuss charge equilibration at the 
Fermi energies, in a kind of overdamped regime. The mechanism is now based 
on the isospin diffusion, on a time scale again ruled by the symmetry term 
at sub-saturation density.

\section{Probing the symmetry energy in nuclear reactions in the Fermi energy
domain}

Heavy ion collisions ($HIC$) in the Fermi energy domain (30-100 AMeV) are 
dominated by fragmentation
mechanisms.  After moderate compression effects during the first stage 
of the collision, the
composite nuclear system expands and eventually separates into fragments, 
whose multiplicity and
characteristics depend on the centrality of the reaction.  
However, as we will illustrate in the following, regardless of the possible
different outcomes, fragment properties are always closely linked to the 
development of density gradients 
along the reaction path. In the energy range considered, the properties 
of nuclear matter below normal
density are mostly concerned, allowing one to investigate the low density 
behavior of the symmetry energy.

In central collisions, where the full disassembly of 
the system
into many intermediate mass fragments (IMF) and particles is observed, 
one can study specifically properties of 
liquid-gas phase transitions occurring in asymmetric matter in the presence
of radial flow. 
In neutron-rich matter in the coexistence region one expects to observe 
 isospin distillation:
fragments (liquid) appear more symmetric
with respect to the initial matter, while light particles (gas) are 
more neutron-rich \cite{baranPR,chomazPR,mue95,bao197,BaranPRL86,margchom03}. 
The  magnitude of this effect
depends on characteristic properties of the isovector sector of the EOS
namely on the value and the slope of the symmetry 
energy at low density 
\cite{baranPR,WCI_betty}.  
Moreover, interesting correlations between fragment isotopic content and 
collective
velocity can be evidenced, also sensitive to the symmetry energy. 

Increasing the impact parameter, fragmentation of the neck region, that is
the overlap region between the two colliding nuclei, takes place.
In this situation, corresponding to mid-peripheral events, 
fragment kinematical properties may keep a memory of the entrance 
channel. The presence of a density gradient between the dilute neck region
and the remaining matter, projectile- and target-like fragments (PLF-TLF),
 induces
interesting isospin effects: The IMF's 
emerging from the neck region are
more neutron rich and this 'isospin migration' mechanism is sensitive to the
symmetry energy.    
For more peripheral events one essentially observes a binary outcome, 
with the
possibility of a dynamically induced fast fission, still triggered by 
the dynamics
of the overlap zone. In reactions between systems with different initial N/Z,
isospin diffusion may occur through the low density interface, governed by  the
strength of the symmetry energy below saturation. 

From one side, these arguments 
lead us to single out 
the isospin signal as a good tracer of the reaction
mechanism \cite{rami}. In fact the new features appearing in neutron rich 
matter
can be used to validate our current interpretation of the  fragmentation 
mechanisms. 
One the other hand, 
isotopic properties and new correlations, once confronted with 
experimental data, can be used
to probe the symmetry term of the EoS at sub-saturation density. 

\subsection{Low Density Behavior of $C_{sym}$ : Isospin Diffusion}

In this subsection we focus on the mechanisms connected to isospin 
transport in 
binary events at Fermi energies. We consider semi-peripheral reactions
between systems having different N/Z and we investigate the diffusion
of the initial N/Z gradient under the conditions dictated by the underlying
dynamical evolution.  This process involves nucleon exchange 
through the low density neck region and hence it is sensitive to
the low density behavior of $C_{sym}$, i.e. of the potential part of the
symmetry energy, \cite{tsang04,rizzo08,sherry}.

\begin{figure} [h]
\vskip 0.2cm
\centering
\includegraphics[width=8.cm]{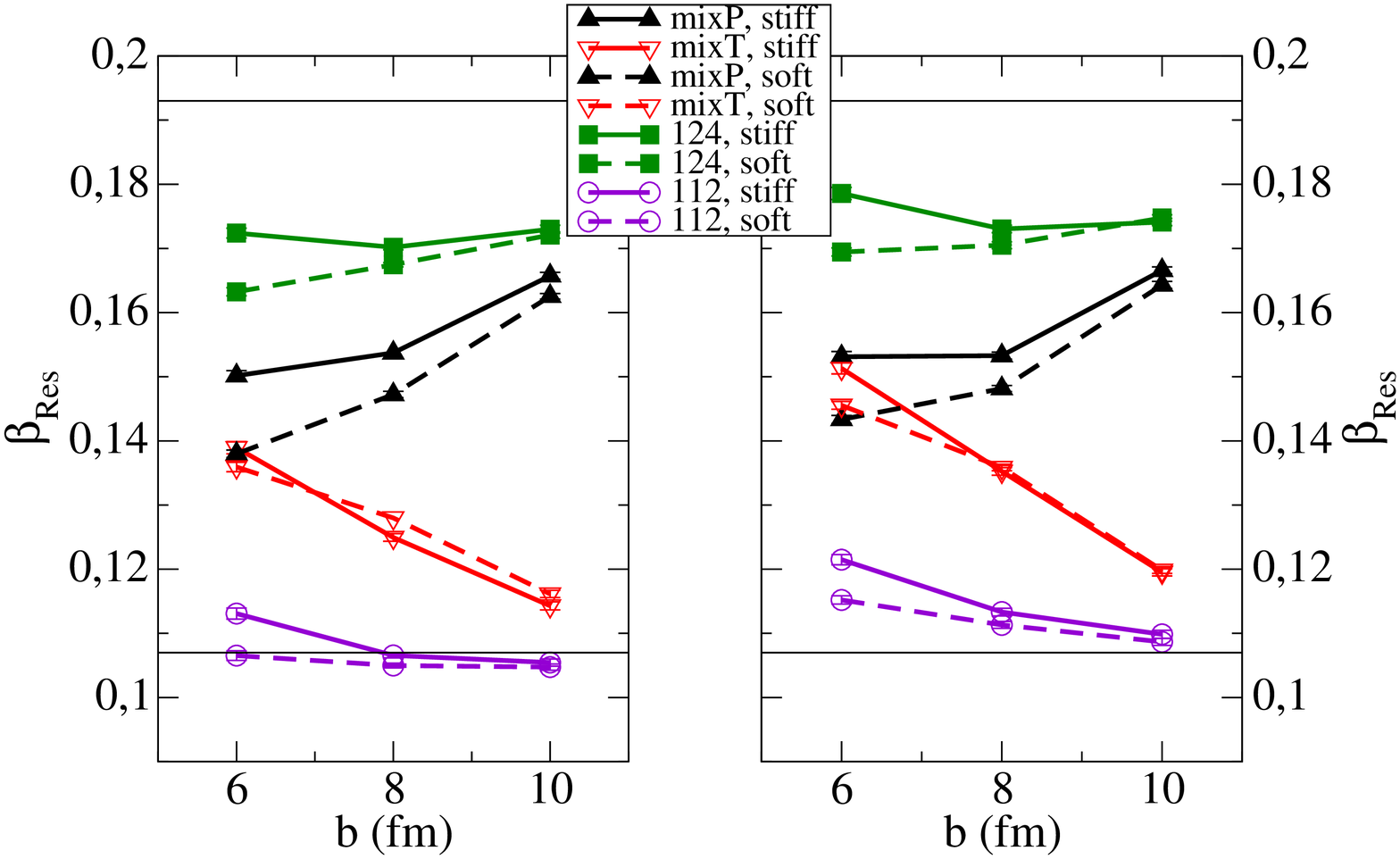}
\caption{Asymmetries of the residues in $Sn+Sn$ collisions at incident 
energies of 
$E=50 AMeV$ (left) and $35 AMeV$ (right) for MD interactions.
 Primary fragments at the $freeze-out$.}
\label{res3550}
\end{figure}

\subsubsection{Asymmetries of  Reaction Components }

As a  direct consequence of the fact that isospin transport takes place, 
projectile-like (PLF) and target-like (TLF) fragments, i.e. the sizeable  
fragments emerging from (mid-)peripheral reactions, 
have different asymmetries with respect to the initial conditions.   
This will be illustrated in the case of
different $Sn+Sn$ ($N/Z=1.48~and~1.24$) reactions ($^{124}Sn +^{124}Sn$ (HH),
 $^{112}Sn +^{112}Sn$ (LL) and the mixed $^{124}Sn +^{112}Sn$ (HL)),
for the impact parameters 
$b=6,8,$ and $10 fm$ and for two incident energies, $35$ and $50$ AMeV. 
This investigation is undertaken
performing SMF simulations, with a momentum dependent (MD) isoscalar
effective interaction (GBD) \cite{rizzo08}. 
To check whether the isoscalar sector of the interaction may
influence isospin effects, we also consider calculations without momentum
dependence. In the latter case we employ a momentum independent (MI) Skyrme 
interaction
having the same compressibility modulus as the GBD interaction. 
As far as the iso-EOS is concerned, calculations are carried out for
two symmetry energy parametrizations: Asysoft and Asystiff.

Results are reported in Fig.\ref{res3550}, for the MD interaction, 
the two Iso-EoS and
the two incident energies. 
It is observed that the asymmetry of the residues  for the mixed $HL$ system
decreases for the   $n$-rich (PLF) and increases for the $n$-poor (TLF) 
partner with respect
to the initial asymmetries, as expected for isospin equilibration.
In the case of the $HH$ and $LL$ symmetric collisions, we cannot have
isospin 
transport and the only variations come from nucleon emissions, 
that leads to a reduction of the
initial asymmetry.
The impact parameter dependence clearly shows  that Iso-EoS effects are more
relevant for more dissipative collisions, i.e. for smaller impact 
parameters and
thus for longer interaction times.  
With respect to different Iso-EoS's, we notice that pre-equilibrium 
emission is more asymmetric
for the soft Iso-EoS, which is expected because of the higher symmetry 
energy and thus 
the larger 
neutron repulsion.

Comparing the results obtained at the two incident energies, 50 $AMeV$ (left) 
and 35 $AMeV$ (right panel), one can see that  
the neutron emission to the 
continuum is 
lower at the lower energy, leading to more initial asymmetric residues.
On the 
other hand the interaction time is larger in this case, leading to more 
equilibration 
in the mixed  system.

\subsubsection{Imbalance Ratio}
Within a first order approximation of the transport dynamics, the relaxation
of a given observable $x$ towards its equilibrium value can be expressed as:
$x_{P,T}(t) - x^{eq} = (x^{P,T} -  x^{eq})~e^{-t/\tau}$, 
where $x^{P,T}$ is the $x$ value for the projectile (P) or the target 
(T) before the diffusion 
takes place, 
$x_{eq} = (x^P + x^T)/2$ is the full equilibrium value, $t$ is the elapsed time
and $\tau$ is the relaxation time, that depends on the 
mechanism under study.   
The degree of isospin equilibration reached in the collision can be 
inferred by looking 
at isospin dependent observables in the exit channel, such as the 
asymmetries of
PLF and TLF, as discussed above.
It is rather convenient to construct the so-called imbalance ratio 
\cite{tsang04,rami,imbalance}:
\begin{equation}
R^x_{P,T} = {(x_{P,T}-x^{eq})} / {|x^{P,T}-x^{eq}|} 
\label{imb_rat}
\end{equation}
 Clearly, this observable measures the difference between the actual 
asymmetry of 
PLF (or TLF) and the full equilibrium value, normalized to the initial
distance (i.e. to the conditions before the diffusion process has started).
In the calculations the latter can be evaluated by looking at the asymmetries
of PLF ( or TLF), as obtained in the symmetric reactions HH and LL 
(where diffusion
does not take place), after
pre-equilibrium nucleon emission is over, see Fig.\ref{res3550}.
The ratio $R$ is $\pm1$ in
the projectile
and target regions, respectively, for complete transparency, 
while it is zero for full equilibration.

\begin{figure}[t]
\centering
\includegraphics[width=7.0cm]{probe8a.eps}
\includegraphics[width=7.5cm]{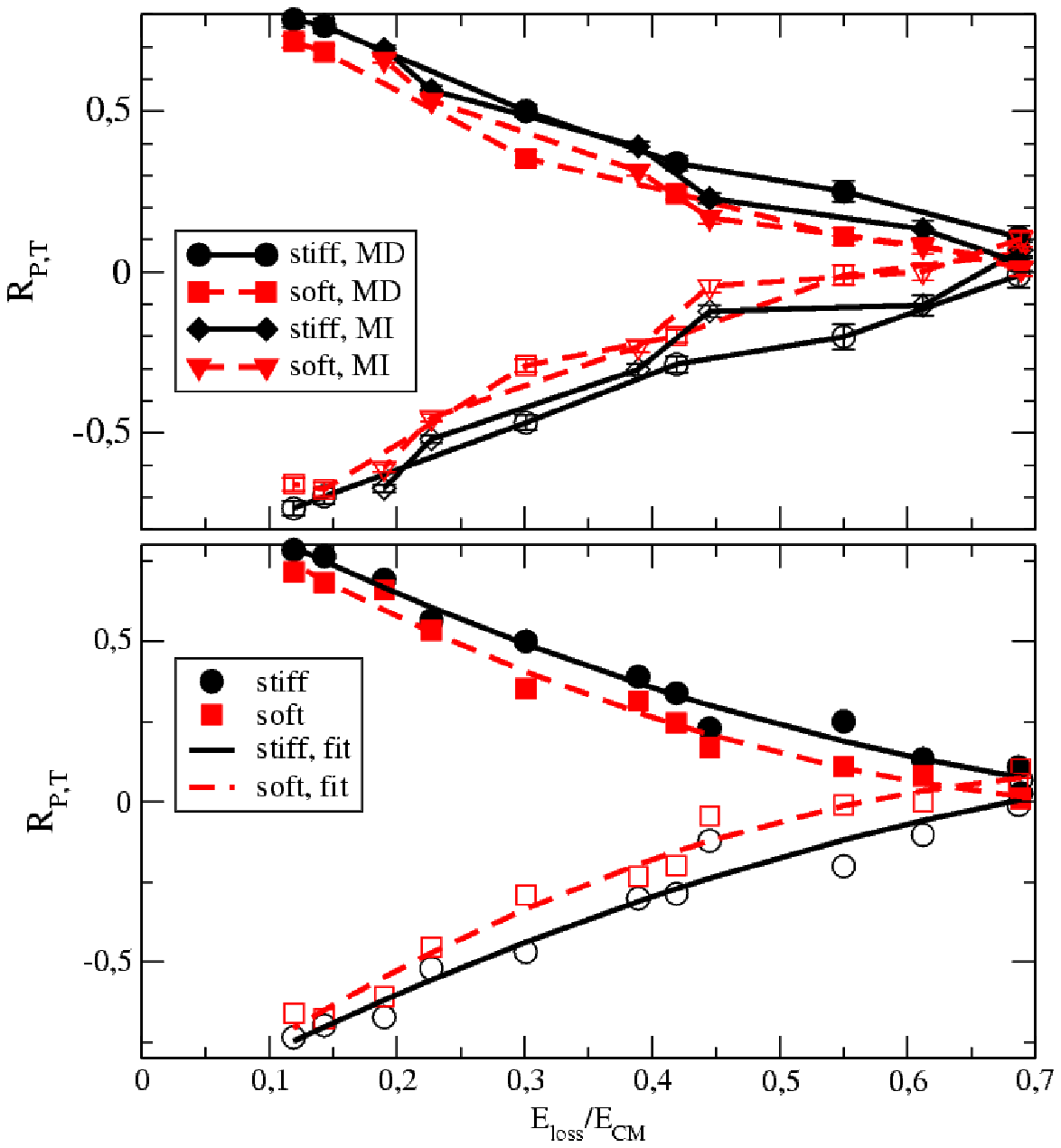}
\caption{ Left Panel.Imbalance ratios for $Sn + Sn$ collisions for 
incident energies
of 50 (left) 
and 35 $AMeV$ (right) as a function of the impact parameter. Signatures of 
the curves: 
iso-EoS stiff (solid lines), soft (dashed lines); MD interaction (circles),
 MI interaction (squares); projectile rapidity ( full symbols, upper curves ),
 target rapidity ( open symbols, lower curves ).
Right Panel. Imbalance ratios as a function of relative energy loss for both 
beam energies. 
Upper: Separately for 
stiff (solid) and soft (dashed) iso-EoS, and for MD 
(circles and squares) and MI 
(diamonds and triangles) interactions, in the projectile region (full symbols)
 and the target region 
(open symbols).
Lower: Quadratic fit to all points for the stiff (solid), resp.
 soft (dashed) 
iso-EoS.}
\label{imb_eloss}
\end{figure}

Within our approximation, the imbalance ratio simply reads 
$R_{P,T} = \pm e^{-t/\tau}$ and  
is actually
independent of the initial asymmetry distance between the reaction partners.
Hence it
isolates the effects of the isodiffusion mechanism, whose strength  is 
determined by the 
relaxation time $\tau$, related to the symmetry energy.
However the degree of equilibration reached in the reaction crucially 
depends also
on the  contact time $t$, i.e. on the reaction centrality. 

From the asymmetries shown before (Fig.\ref{res3550}) one can 
build the imbalance 
ratios,
 according to Eq.(\ref{imb_rat}), that are shown  
in Fig.\ref{imb_eloss}, Left Panels, for MD calculations (circles),
for the incident energy of  50 $AMeV$ (left) and 35 $AMeV$  
(right).
Results for 
stiff and
soft Iso-EoS are represented with
 solid and dashed lines, and  those  for the projectile and target 
rapidity  regions
 with full and open
 symbols  (upper and lower curves) , respectively.
We show also results obtained in the case of the MI interaction (squares).  

The
 equilibration is 
larger (R smaller) both for lower energies  (right)  and for MI 
interactions
 (dashed curves).  In fact in both cases the reaction is slower and 
thus the interaction
 time longer, 
leading to more equilibration; for the lower energy because of the slower 
speed, and for the MI
 interaction because of the less repulsive isoscalar mean field. In comparing
 the two 
Iso-EoS's we see that the equilibration is larger for the soft Iso-EoS, since 
the higher 
symmetry energy leads to a larger diffusion contribution to the 
isospin current. 

To isolate isospin effects, 
one could study the asymmetry imbalance ratio as a function of 
the interaction time
(or of an observable directly related to it).

To this purpose we propose to study the correlations between isospin 
equilibration and
the kinetic energy loss, that is adopted as a selector of the reaction 
centrality and,
hence, of the contact time $t$. 
We define the kinetic energy loss per 
particle as
\begin{equation}
E_{loss}  =  E_{cm} - \frac{E_{kin}}{A_{PLF}+A_{TLF}} \, , \,
E_{cm}  =  \frac{E_{lab}}{A_P} \frac{A_P A_T}{(A_P+A_T)^2} \, ,
\label{E_loss}
\end{equation}
where $A_P, A_T, A_{PLF}, A_{TLF}$ are the masses of the initial projectile 
and target, and of 
the final projectile-like and target-like fragments, respectively. Here 
$E_{cm}$ is the initial 
energy, per nucleon,  available in the $cm$ system. $E_{kin}$ 
is the final total kinetic 
energies of the fragments in the $cm$ system.  
We mention that the study of isospin equilibration 
as a function of
the heavy residue excitation energy (related to the kinetic energy loss  )
was suggested in Ref. \cite{soul04}.
In Fig.\ref{imb_eloss}, Right Panels, we report the correlation between
$R_{P,T}$ 
and the total kinetic energy loss (normalized to $E_{cm}$) for the full 
set of calculations performed.  
On the bottom part of the figure, where all results are collected together, 
one can see
that all the points essentially follow a given line,
depending only on the symmetry energy parametrization adopted. A larger
equilibration (smaller $R$) is observed in the $Asysoft$ case, corresponding to
the larger value of $C_{sym}$.
We mention that,
according to its definition, the imbalance ratio does not change if one 
considers
as observable $x$, instead of the asymmetry of PLF and TLF,
other observables  linearly correlated to it and more accessible from 
the experimental
point of view, such as isoscaling coefficients, ratios of 
production of light isobars \cite{WCI_betty} or isotopic
content of light particle emission \cite{Gali}.

\begin{figure}[htbp]
\centering
\resizebox{0.5\textwidth}{!}{%
\includegraphics{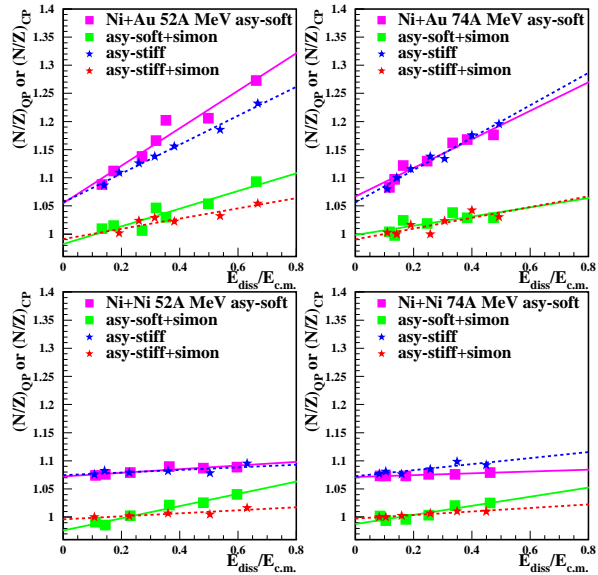}}
\caption{(color online) Isospin ratio of the quasi-projectile vs 
dissipated kinetic energy, 
for the two reactions and the two energies. For the stiff calculation, black 
stars and dotted lines display (N/Z)$_{\mathrm{QP}}$ and grey stars and dotted
lines the (N/Z)$_{\mathrm{CP}}$ (BNV calculation followed by SIMON). Same
conventions  for the asy-soft case displayed by squares and full lines. The
lines correspond to linear fits. Taken from ref.\cite{Gali}.}
\label{figure1_emma}
\end{figure}

\subsubsection{Comparison with Experimental Data}
An experimental study of isospin diffusion as a function of the dissipated
kinetic energy  has been recently performed by the Indra collaboration 
\cite{Gali}. 
 
Two systems, with the same projectile, $^{58}Ni$,
and two different targets ($^{58}Ni$ and $^{197}Au$), at incident 
energies of 52 AMeV and
74 AMeV have been considered. 
This choice gives access to isospin effects in 
different conditions of 
charge (and mass) asymmetry, with respect to the analysis discussed in 
the previous
sections,  and to their evolution with the energy
deposited into the system. In the symmetric Ni + Ni system isospin effects are
essentially due to the pre-equilibrium emission. On the contrary, in the 
charge (and
mass) asymmetric reactions, one can observe isospin transport between the two
partners, leading to the neutron enrichment of the PLF.
Hence, in the following, we will focus on the PLF properties.  
An isospin-dependent variable, correlated
to the PLF asymmetry, is constructed 
from the isotopically identifed particles
emitted from the PLF:
\begin{equation}
                  (< N > / < Z >)_{CP} =  \sum_{N_{ev}} \sum_\nu N_\nu /  
\sum_{N_{ev}} \sum_\nu   P_\nu,
\end{equation}

where $N_\nu$ and $P_\nu$ are respectively the numbers of neutrons and 
protons bound in the
particle $\nu$ ,  $\nu$ being d, t, $^3$He, $^4$He, $^6$He, $^6$Li, 
$^7$Li, $^8$Li, $^9$Li, $^7$Be, 
$^9$Be, $^{10}$Be; since neutrons are not measured free
protons are excluded. $N_{ev}$ is the number of events contained in 
a given bin of dissipated energy (or kinetic energy loss), that is used as 
a selector of the
reaction centrality.

Simulations have been carried out considering two 
parametrizations of
the symmetry energy: Asy-soft and Asy-stiff.  
Results are presented in Fig.\ref{figure1_emma}, that shows
the evolution of the N/Z ratio of the PLF, $N/Z_{QP}$,  
as a function of $E_{diss}/ E_{c.m.}$
One can see that N/Z increases with the centrality of the collision for the two
systems and the two beam energies. For the Ni+Ni system the variation of 
N/Z with
centrality is small, and attributed to pre-equilibrium emission. 
Indeed, for this system with a small neutron excess, more protons
are emitted during the pre-equilibrium stage, especially in the 
asy-stiff case, 
due to the coupled effect of Coulomb
repulsion and a less attractive symmetry potential for protons. This 
effect increases
with the incident energy. On the contrary, the Asy-soft case tends to emit more
preequilibrium neutrons leading to a lower N/Z ratio \cite{baranPR}.
    
The evolution with centrality is much more pronounced for the 
neutron-rich and
asymmetric Ni+Au system. In addition to pre-equilibrium effects, isospin 
transport
takes place between the two partners of the collision, and increases with 
the violence
of the collision, more for Asy-soft.
It must be underlined that isospin
equilibration is nearly reached, for the most central collision, at the 
lower energy for the Asy-soft EoS.
However, when comparing diffusion effects corresponding to different 
Iso-EoS parametrizations, 
one should keep in mind that the isotopic content of the pre-equilibrium 
emission is 
also dependent on the Iso-EoS.
In the Asysoft case, for instance, isospin diffusion is more effective but, 
at the same time,
more neutrons are removed from the system by fast nucleon emission. In the 
mixed reaction, this 
reduces 
the sensitivity of the PLF asymmetry to the Iso-EoS. The same arguments 
hold for the comparison
of results obtained at two different beam energies. In fact, 
pre-equilibrium emission is
more abundant at higher energy. 
Isospin transport effects would be isolated  
by constructing imbalance
ratios, as discussed before. However, such analysis 
is not possible for this set of data. 

\begin{figure}[htbp]
\centering
\resizebox{0.5\textwidth}{!}{%
\includegraphics{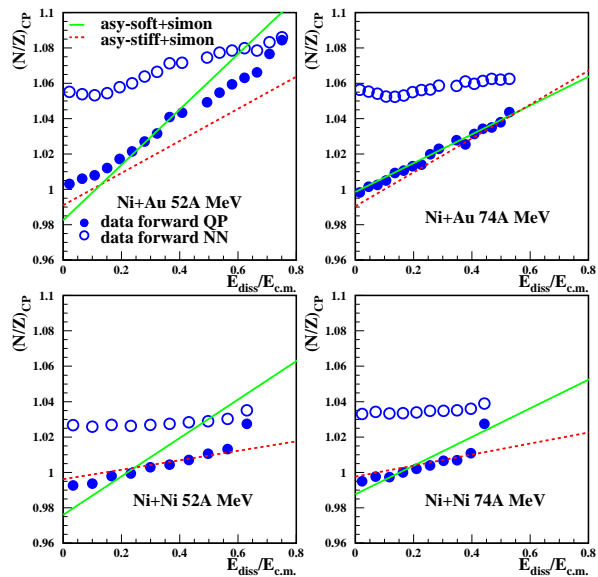}}
\caption{(color online) Isospin ratio of complex particles for Ni
 quasi-projectile vs dissipated kinetic energy, for the two reactions and 
the two energies.
Circles correspond to the experimental data, open for data forward of the 
N-N velocity~\cite{galichet}, full for data forward in the QP frame.
Dotted lines and full lines as in fig.~\ref{figure1_emma}. Adapted from
\cite{Gali}.}
\label{figure2_emma} 
\end{figure}

In  figure \ref{figure1_emma} are also plotted the results concerning 
the variable $(N/Z)_{CP}$ , calculated
after de-exciting the hot primary PLF's with the help of the SIMON 
code \cite{Charity}. 
The
values of $(N/Z)_{CP}$ are always smaller and the evolution with 
dissipation is generally
flatter than that of the N/Z of the primary PLF: secondary decay weakens 
the isospin effects. 
However, $(N/Z)_{CP}$ , which appears in the simulations to be linearly 
correlated with the
N/Z of the primary PLF, is thus a good indicator of isospin transport 
effects and is
sensitive to the Iso-EoS.
Moreover, this observable can be directly compared to experimental data. 
This is shown
in Fig.\ref{figure2_emma}.  Open points show the values obtained 
forward in the 
nucleon-nucleon (NN) frame.
In this case mid-rapidity particles and those coming from the
PLF de-excitation are mixed up. 
Full points in Fig.\ref{figure2_emma}  are related to 
the values of $(N/Z)_{CP}$ forward in the
PLF frame.  
They are representative of the isotopic 
content of the particles emitted
from the PLF only and can be compared with 
the results of the simulation, displayed in 
Fig.\ref{figure2_emma} by the lines. 
When looking globally at the results for the four cases treated
here, the agreement is better when the Asy-stiff EoS is used, 
i.e. a linear increase
of the potential term of the symmetry energy around normal density.  
This is in reasonable agreement with the 
conclusion drawn from isospin diffusion
in Sn+Sn systems at 50 AMeV,\cite{bettynew} see before, 
and matches the 
one derived from the competition
between dissipative mechanisms for Ca+Ca,Ti at 25 AMeV \cite{Chimeranew}.

\subsection{Isospin Distillation in Presence of a Radial Flow}

The study of isospin effects in central collisions and, in particular, 
of the isospin
distillation mechanism is interesting  also in a more general context:
in heavy ion collisions the dilute phase appears during the expansion
of the interacting matter. Thus one can investigate effects of the coupling 
of expansion, fragmentation and distillation
 in a two-component (neutron-proton) system \cite{col08}.
In the following we will discuss correlations
between the isotopic content  of IMF's
and kinematical properties in central multifragmentation reactions. 
Fragmentation originates from the break-up of a composite source
that expands with a given velocity field.  
Since neutrons and protons experience different forces, 
one may expect a different radial flow for the two species. Being these 
forces connected 
to the density dependence of the symmetry energy, one should be able to 
extract 
information on its properties. 

Such analysis is, in a way, complementary to the study of the isospin content 
of pre-equilibrium nucleon emission \cite{Famiano,Fam1}, 
because it looks at the same phenomenon 
from the point of view of the fragmenting residual system. In fact, 
we will see that the 
behaviour is often the opposite of the one observed for the 
pre-equilibrium emission.

\subsubsection{Fragmentation Path in Central Collisions}
We will focus on central collisions, $b = 2~fm$, in symmetric reactions
between systems having three different initial asymmetries: 
$^{112}Sn + ^{112}Sn, ^{124}Sn + ^{124}Sn,$ and
$^{132}Sn + ^{132}Sn$ with $(N/Z)_{in}$ = 1.24, 1.48 and 1.64, respectively. 
The considered beam energy is 50 MeV/nucleon.
Calculations are carried out for two parameterizations of $E_{sym}$, 
Asysoft and
Asystiff. 
The first two reactions, $^{112}Sn + ^{112}Sn$ and $^{124}Sn + ^{124}Sn$,
have been widely  investigated both from the experimental 
and theoretical point of view \cite{bar02,tsangprl1,bao,liu}. 

We first illustrate some general properties of the fragmentation mechanism 
in central events, 
as described by the SMF model.  
After the collisional shock and the initial compression, the composite
nuclear source expands. Along this expansion, small density fluctuations are
amplified by the unstable mean-field and large amplitude density gradients are
developed. This process ends up with the formation of several fragments,
corresponding to the high density bumps, located on 
a bubble-like configuration 
\cite{chomazPR,Fabbri}. 
Their average multiplicity is approximately equal to 6 for the reactions 
considered here \cite{bar02}.
Several nucleons are emitted, prior to fragmentation, at the early stage
(pre-equilibrium emission) and/or are evaporated while fragments are formed. 
Primary fragments are identified by applying 
a coalescence procedure to the matter with density larger than 
$\rho_{cut} = 1/5~\rho_0$ (that we classify as 'liquid phase').
The remaining nucleons are considered as belonging to the 'gas phase'.

\begin{figure}
\centering
\vspace{1.3cm}
\includegraphics[width=7.5cm]{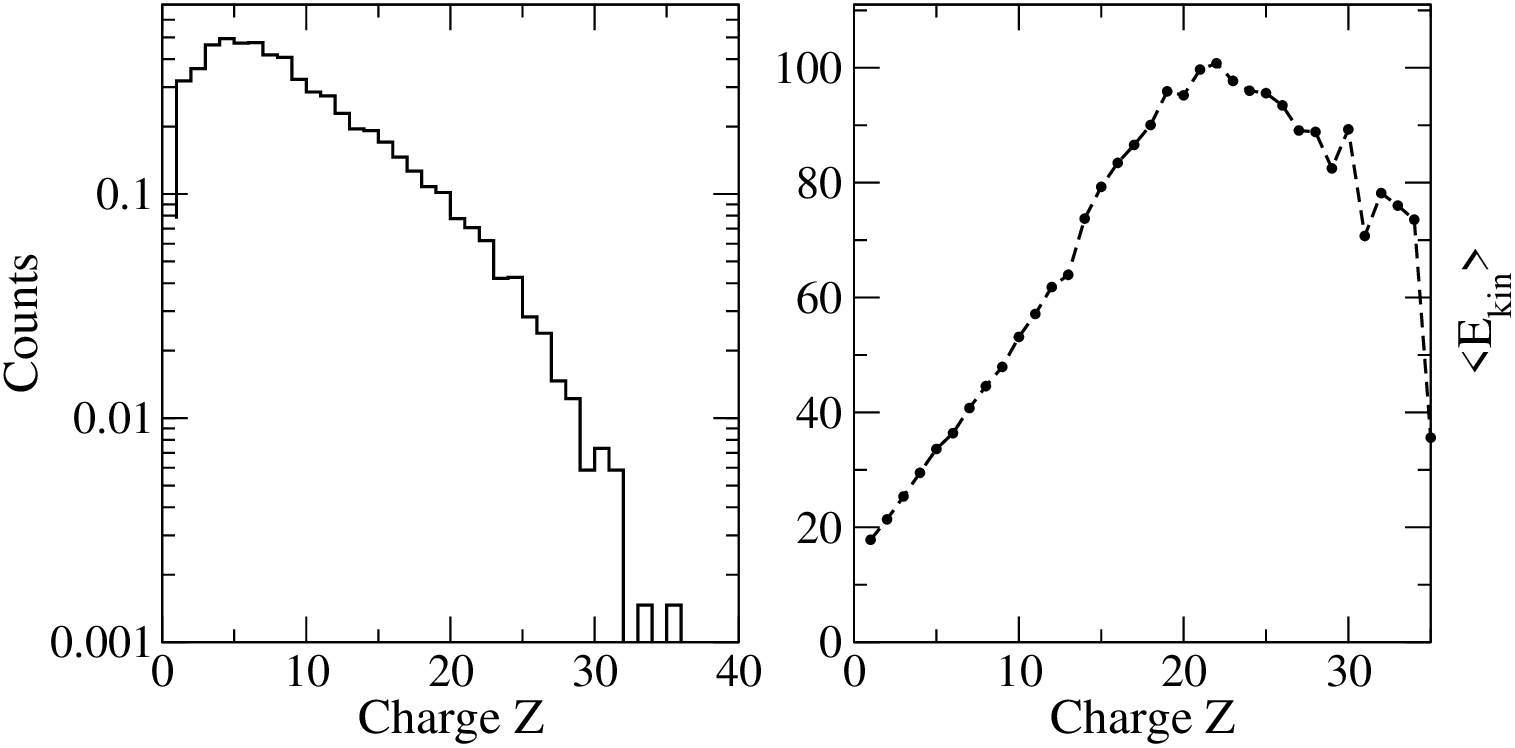}
\hspace*{5pt}
\includegraphics[width=7.0cm]{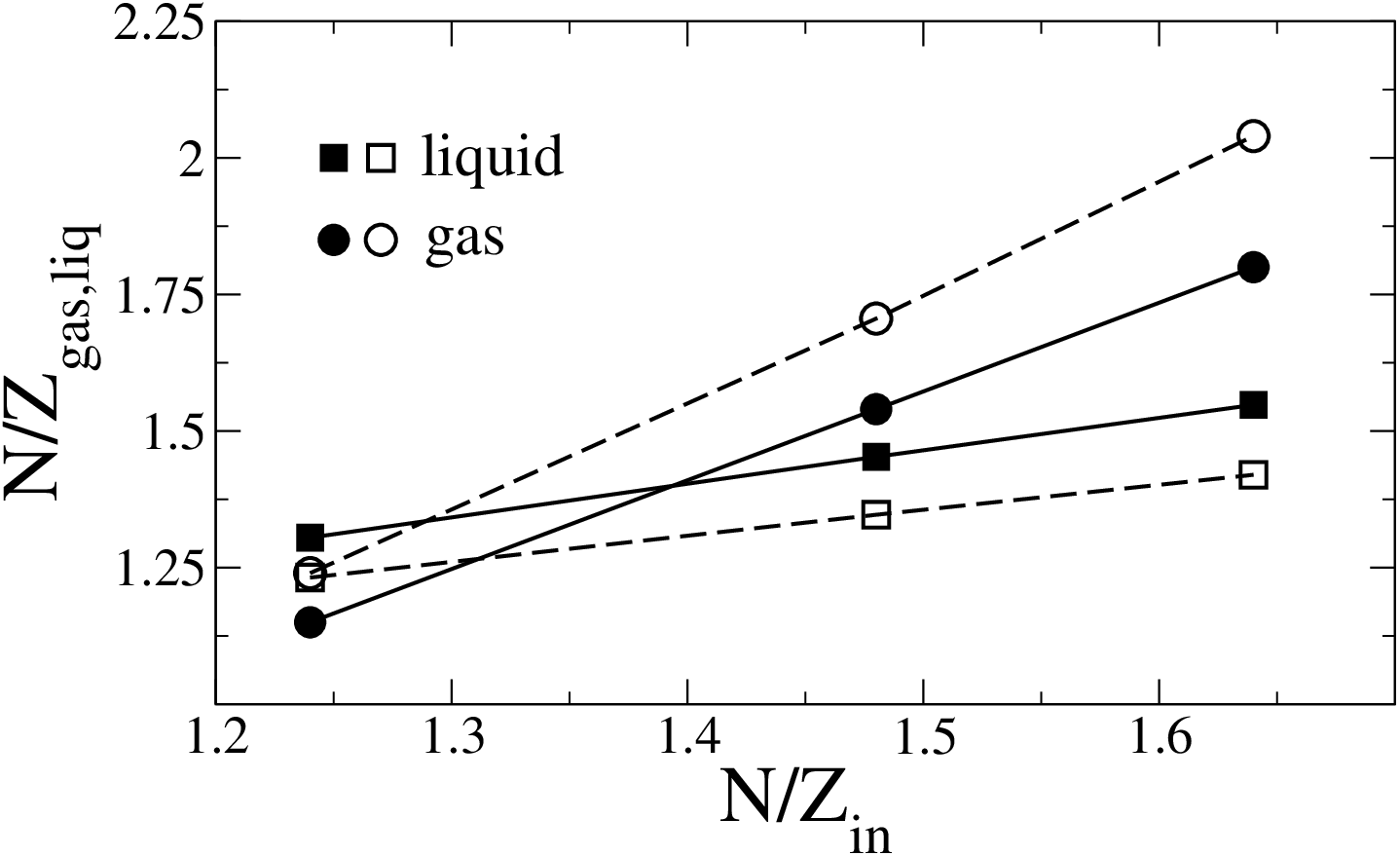}
\caption{Left Panel: Fragment properties in the reaction $^{112}Sn + ^{112}Sn$ 
at b = 2 fm, E/A = 50 MeV/nucleon, t = $300~fm/c$. 
Left: charge distribution. Right: average kinetic energies.\\
Right Panel: Asymmetry $N/Z$ of the gas (circles) and of the liquid (squares) 
phase for central $Sn + Sn$ collisions with different initial $N/Z$.
Full lines and full symbols refer to the Asystiff, dashed
lines and open symbols to the Asysoft parameterization. 
}
\label{chargeasy}
\end{figure}

In Fig.\ref{chargeasy} (Left Panel) we show the fragment charge 
distribution (left) and the average kinetic
energy as a function of the charge Z (right). From the linear increase of 
the kinetic energy with the fragment charge Z one can estimate the radial 
collective flow.
The change of trend observed for big fragments reflects their different 
formation
mechanism, via recombination effects or from PLF/TLF residues.
We restrict our analysis to fragments with charge between 3 and 10 (IMF)
produced in the fast break-up of the system; they represent the
$liquid~phase$ with a clear evidence of radial flow.

The average $N/Z$ of emitted nucleons (gas phase) and of the IMF's
is presented in Fig.\ref{chargeasy} (Right Panel) as a function of the initial 
asymmetry, $(N/Z)_{in}$, 
of the three colliding $Sn$ systems.

Generally, the gas phase is seen to be more neutron-rich while the IMF's 
are more symmetric. 
This is due to the combined action of the pre-equilibrium emission, that 
reduces
the neutron excess of the composite system, and of the distillation mechanism
acting in a later stage, while fragments are formed. 

This trend is stronger in the
Asysoft relative to the Asystiff 
case, since the symmetry energy is larger below saturation 
in the former case \cite{bar02}.
The difference between the asymmetries of the gas and liquid phases 
increases
with the $(N/Z)_{in}$ of the system, and is always larger in the 
Asysoft case.
It should be noticed that the isotopic 
content of the gas phase appears more sensitive to the Iso-EoS
than the asymmetry of the fragments. As one can see from
Fig.\ref{chargeasy} (Right Panel), for the IMF's the difference between 
the two 
EoS's is just about $8\%$.

\begin{figure}
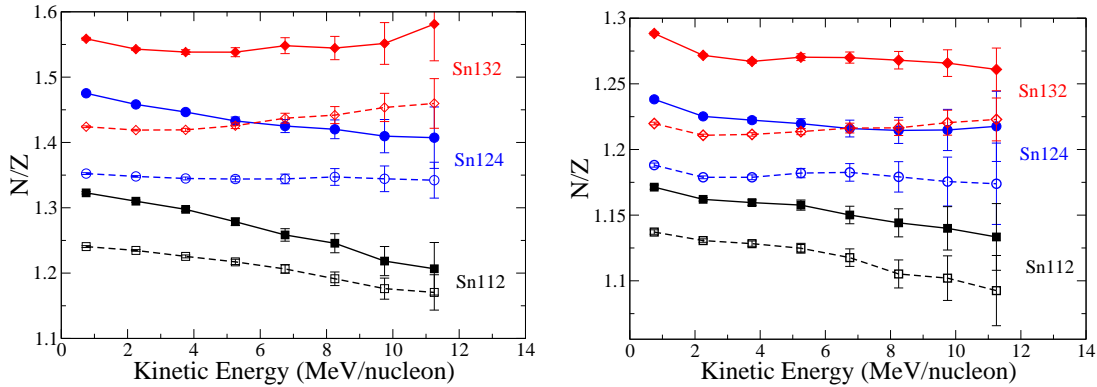

\centering
\vspace{0.8cm}
\includegraphics[width=7.cm]{probe12a.eps}
\hspace*{5pt}
\includegraphics[width=7.cm]{probe12b.eps}
\caption{Left Panel.
Fragment asymmetry $N/Z$ (see text) as a function of the
 kinetic energy for different mass symmetric $Sn + Sn$ collisions at b=2 fm, 
E/A = 50 MeV/nucleon. 
Solid lines are for the asystiff and dashed lines for the asysoft symmetry 
energy, and the different symbols distinguish the different collision 
systems.\\
Right Panel.
Final fragment asymmetry $N/Z$, i.e. after evaporation,  as a
 function of the 
kinetic energy (same format as in the Left Panel). 
} 
\label{nzkin}
\end{figure}

\subsubsection{Isospin-Velocity Correlations}
Within a full detection of fragment properties, the investigation of
the possible existence of correlations between size, isotopic and
kinematical observables brings new information about the fragmentation
mechanism, the expansion and cooling dynamics, and the nuclear interaction 
in the low density regime. 
For the range of charges considered in our analysis, $3<Z<10$, the average 
fragment size is loosely correlated to the velocity.
In fact, as one can see from Fig.\ref{chargeasy} (Left Panel), small 
fragments have nearly
the same collective velocity (kinetic energies proportional to the masses). 
Now we discuss more in detail the correlations between fragment isotopic 
content and kinematic properties. 
As a measure of the isotopic composition of the IMF's we will consider
the sums of neutrons, $N = \sum_i N_i$, and protons, $Z = \sum_i Z_i$, 
of all IMF's in a given kinetic energy bin (here taken as 
1.5 MeV/nucleon), in each event, taken at $t=300fm/c$ (primary fragments). 
Then we take the ratio $N/Z$ and consider
the average over the ensemble of events. 
In Fig.\ref{nzkin} (Left Panel) we report the dependence of this 
fragment asymmetry on 
the kinetic
energy for the three reactions, and for the two parametrizations of the 
symmetry energy. 
The magnitude of the fragment asymmetry changes with the symmetry energy,
as expected.  
Moreover the slopes of the curves in Fig. \ref{nzkin} (Left)
 appear 
particularly characteristic of the asymmetry of the initial system and of the 
stiffness of the symmetry energy. In fact, we expect two opposing trends since
the Coulomb energy alone will accelerate the more proton-rich fragments. 
Thus the slope of the curve $N/Z(E_{kin})$ will be negative, and more so for 
more proton-rich systems (e.g. in Fig.\ref{nzkin}(Left) the system 
$^{112}Sn + ^{112}Sn$ has the largest 
negative slope). The symmetry energy, on the other hand, is more repulsive for 
more neutron-rich fragments, and thus the slope 
should be positive, more so for the softer symmetry energy 
(e.g. in Fig.\ref{nzkin}(Left) the 
slope for asysoft is always larger than for asystiff). When both forces are
 present there 
is a compensation between these two trends, and the final slope can be both 
negative or positive. It should increase for an Asysoft symmetry energy 
and neutron-richer systems. This is observed in Fig.\ref{nzkin}(Left) as a 
general trend.

\subsubsection{Secondary Decay Effects}

So far we have discussed features related to primary fragments. However, 
in order
to compare with experimental data, one cannot avoid treating the 
de-excitation process.
For this we have calculated the excitation energy of the 
fragments in each event (which is found to be about $2.5 \pm 1$ MeV/nucleon on 
the average), 
and used the statistical evaporation code  SIMON \cite{Charity} for the decay. 
The fragment $N/Z$ ratios for the secondary fragments are shown in 
Fig.\ref{nzkin}(Right Panel)
in the same format as in Fig.\ref{nzkin}(Left Panel) for the primary fragments.
The final fragment $N/Z$ ratio is found to be reduced by the secondary decay, 
due to the abundant 
neutron evaporation. It approximately follows the relation 
$(N/Z)_{fin} \approx a(N/Z) + b$, with $a$ and $b$ 
not much depending on the
fragment kinetic energy and initial asymmetry ($a \approx 0.55$ and 
$b \approx 0.45$).
It is also seen that a linear energy dependence is less well fulfilled than 
for the primary fragments, 
in particular for low kinetic energies.
However, one can still appreciate the evolution of the $N/Z$ energy dependence 
with the neutron
richness of the initial systems and the difference between the predictions of 
the two iso-EOS's,
though the sensitivity is reduced with respect to 
primary fragments (compare Left and Right Panels in Fig.\ref{nzkin} ).
Finally
we mention that, as far as the final fragment $N/Z$ is concerned, 
results may depend on the evaporation code considered. This introduces an
additional degree of uncertainty in the comparison to experimental data.
The issue of a more accurate treatment of the de-excitation chain of
 neutron-rich
nuclei should be critically addressed in the future.

\subsection{Isospin Dynamics in Neck Fragmentation}

 \subsubsection{Experimental Survey}

It is now quite well established that the largest part of the reaction
cross section for dissipative collisions at Fermi energies goes
through the {\it Neck Fragmentation} channel, with $IMF$s directly
produced in the interacting zone in semiperipheral collisions on very short
time scales \cite{baran2004,wcineck,colonnaNPA589}. We can predict 
interesting 
isospin transport 
effects for this new
fragmentation mechanism since clusters are formed still in a dilute
asymmetric matter but always in contact with the regions of the
projectile-like and target-like remnants almost at normal densities.
In presence of density gradients the isospin transport
is mainly ruled by drift coefficients and so
we expect a larger neutron flow to
 the neck clusters for a stiffer symmetry energy around saturation, 
\cite{baranPR,baranPRC72}. 

  A systematic experimental study of midvelocity emission as a 
function of beam energy,
violence of the collision and mass of the system has been performed
in the last years by several groups
\cite{piantelli2006,piantelli2007,milazzo2005,theriault2006,hudan2005},
using different fragment multidetectors. They all agree on the evidence of 
two distinct sources of fragment production, one at mid-rapidity
(the neck fragmentation) and one at the Quasi-Projectile rapidity
(PLF* evaporation) ( the detection velocity threshold was too high
to clearly see the Quasi-Target emissions). All experiments also agree
on a neutron enrichment of the mid-rapidity fragments, even in the cases
of the same isospin asymmetry of the colliding ions, like  $Ni+Ni$
(almost charge symmetric) 
\cite{milazzo2005}, $^{64}Zn+ ^{64}Zn$ \cite{theriault2006},
$^{93}Nb+ ^{93}Nb$ and  $^{116}Sn+^{116}Sn$ \cite{piantelli2006,piantelli2007}.

   In the case $^{114}Cd+ ^{94}Mo$ at $50$ AMeV  \cite{hudan2005} it was
also shown that the fragments with charge $3 \le  Z \le 20$ at
midrapidity exhibit a rather surprising similar neutron enrichment 
between central 
and midperipheral collisions. This could be an indication of an Asy-stiff
behavior of the symmetry term below saturation which implies a reduced
isospin distillation for fragments produced in central events via spinodal
mechanism (see Sect.4.2) and a larger neutron flow to the neck region for 
semipheripheral
collisions.

 A very accurate analysis of neck fragmentation has been possible with the
use of the CHIMERA Multidetector at
Laboratori Nazionali del Sud-INFN, Catania, that for the very low threshold
characteristic of the telescopes was allowing also a good detection 
of the emissions from the Target region. The reactions
$^{124}Sn+ ^{64}Ni$ and $^{112}Sn+ ^{58}Ni$ at $35$ AMeV were investigated in
inverse kinematics \cite{dynfiss05,russotto2006}.
These conditions allowed an easy distinction between IMFs produced in the  
PLF, TLF 
and ``neck'' sources, allowing also important angular and velocity 
correlation measurements.
A class of events was clearly identified for which the
IMFs show deviations from velocity Viola systematics simultaneously
with respect to PLF and TLF \cite{velcorr04}. 
The angular distribution of those IMFs originating from the neck-like structure
corresponds to a quite aligned configuration. The $\Phi_{plane}$, the angle
between the fission axis (defined by the PLF-IMF system) projected in the 
reaction
plane and scission  axis (defined by PLF-TLF system) is centered around zero.
Such alignment indicates that after the splitting of initial composite
systems the ensemble PLF-IMF will rotate as a whole not for long period,
resulting an early decoupling of the neck region from the two residues.
The $\Phi_{plane}$ distribution of heavier IMF's become wider while their
velocities along beam axis suggest a continuous transition towards
cases when they are more correlated with one of the two heavy residues
and so emitted at later times without preferential alignment.

  From such  kinematic correlations it is possible, using Coulomb 
trajectory
calculations, to calibrate a time-scale for fragment emission. These 
estimates lead
to the conclusion that the lighter IMFs are produced between $40$ fm/c 
to $80$ fm/c
from the reseparation while the heavier ones are formed even at 
$120$ fm/c
or longer. 
We can expect to see very different isospin content of the fragments 
in the different velocity and angular correlation bins:
 for IMFs with a given
charge the greatest isospin value is acquired by those with the 
largest deviations from Viola systematics and the highest degree 
of alignment \cite{defilippo09}.

From these exclusive data we can extract important information on the
$Iso-EoS$ using microscopic $ab~initio$ transport approaches.

\begin{figure}
\centering
\vskip 0.3cm
 \includegraphics[scale=0.26]{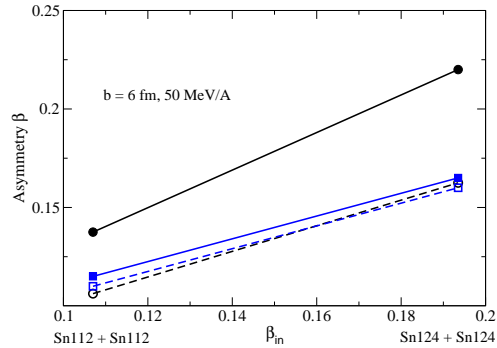}
\caption{Ternary events in semiperipheral $Sn+Sn$ collisions at $50~AMeV$.
Asymmetry of IMF's (circles) and PLF-TLF (squares), as a
function of the
system initial asymmetry, for two Iso-EOS choices: Asystiff (full lines) and
Asysoft (dashed lines).}  
\label{neckasy}
\end{figure}

\subsubsection{Isospin Tracer of the Reaction Mechanism and Symmetry Energy}

  Several transport models were involved to explain various aspects 
of the reaction mechanisms and related isospin dynamics 
\cite{papa2007}, \cite{hudan2006}, \cite{planeta2008}. We employ
Stochastic Mean Field transport simulations, see Sect.2, able to account for 
an accurate description of mean-field dynamics, very important at these 
energies, consistently coupled to dynamical fluctuations, essential to 
account for instabilities and fragment formation.

The first point is to show the sensitivity to the density dependence of the 
symmetry term of the Isospin Migration from the PLF/TLF to the Neck region.
We have analysed 200 SMF ternary events for semiperipheral 
$^{112}Sn+^{112}Sn$ and
$^{124}Sn+^{124}Sn$ collisions at $50~AMeV$.
In Fig.\ref{neckasy} we present the correlation between the average 
asymmetry parameter
$\beta \equiv (N-Z)/A$ of the Fragments emitted at mid-rapidity and of the
PLF/TLF residues vs the initial asymmetry of the colliding ions, using the 
stiff and soft density dependence of the symmetry term below saturation.
As expected from the drift contribution (dominant here since the initial 
distribution of the asymmetry is uniform) proportional to the symmetry 
energy slope below saturation, we see a much larger neutron enrichement 
in the neck fragments for the Asystiff choice, Fig.\ref{esym}, more 
evident in the
$^{124}Sn$ n-rich case.

\begin{figure}
\centering
 \includegraphics[scale=0.30]{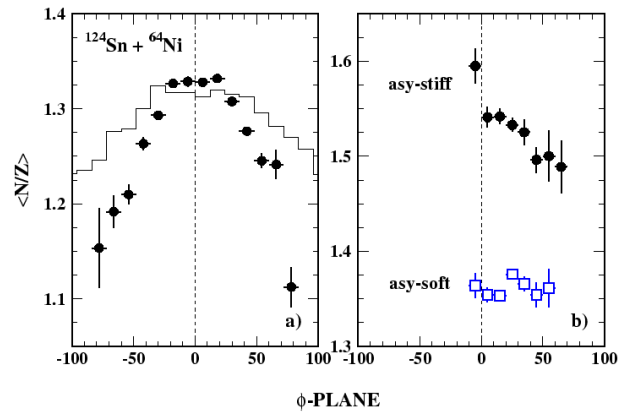}
\caption{Correlation between $N/Z$ of $IMF$ and $alignement$ in ternary
events of  the $^{124}Sn+^{64}Ni$
reaction at $35~AMeV$. $Left~Panel$. Exp. results: points correspond to fast 
formed $IMF$s (Viola-violation selection); histogram for all $IMF$s at 
mid-rapidity (including statistical emissions). $Right~Panel$. Simulation 
results: squares, Asysoft; circles, Asystiff.}  
\label{nzphi}
\end{figure}

A very nice new analysis has been performed on ternary events in the 
$Sn+Ni$ data at $35~AMeV$
by the Chimera Collab.,\cite{defilippo09},
 see Fig.\ref{nzphi} left panel.
For the mid-rapidity IMFs a strong correlation between neutron enrichement, 
alignement and Viola violation (when the 
short emission time selection is enforced) is seen, that can be reproduced 
only with 
a stiff behavior of the symmetry energy, Fig.\ref{nzphi} right panel 
(for primary fragments) \cite{erice08}. A more detailed study for each 
IMF Isotope clearly shows the same effect for the same kinematic selection bins
\cite{defilippo09}.
All that represents a 
clear evidence in favor of a relatively large slope (symmetry pressure) 
around saturation. We note a recent confirmation from structure data,
i.e. from monopole resonances in Sn-isotopes \cite{garg_prl07}.

 In conclusion we can figure out
   a continuous transition from fast produced fragments via 
spinodal and neck instabilities
   to clusters formed in a dynamical fission of the projectile(target) 
   residues up to the evaporated ones (statistical fission). Along this 
   line it would be even possible to disentangle the effects of volume
   and shape instabilities.
From the previous discussion on the 
neutron enrichment of the overlap ("neck") region we can expect that the
IMF Isospin content could be a good tracer of the Reaction Mechanism.

 In the following we suggest new fragment mass-velocity-isospin correlations
particularly sensitive to the various mechanisms as well as to the 
isovector part of the in-medium
nuclear interaction. We remind that a study of just mass-velocity correlations
for the fragmentation of quasiprojectiles
was performed by Colin el al. \cite{colin2003} within the INDRA collaboration,
revealing some evidence of break-up of very elongated structures.

\subsubsection{IMF Mass-Charge-Velocity Correlations} 

\begin{figure}
\begin{center}
\includegraphics*[scale=0.3]{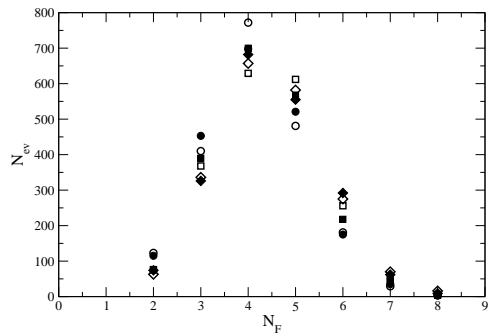}
\end{center}
\caption{Fragment multiplicity distribution at $b=4fm$. Circles 
corresponds to LL,
squares to HH while diamonds are associated with EE entrance channel 
combinations. Full
symbols: Asysoft EoS. Open symbols: Asystiff EoS.}
\label{mult}
\end{figure}

 Here we focus on mass symmetric $Sn+Sn$ reactions at 50 AMeV,
intensively analyzed in the recent years at MSU \cite{lynch2009}.
We present 
   a comparative study of the reactions $^{132}Sn+^{132}Sn$ (EE system),
$^{124}Sn+^{124}Sn$ (HH) and $^{112}Sn+^{112}Sn$ (LL)  at $50MeV/A$.
We shall focus on the value of impact parameter $b=4 fm$ where 
some memory of the entrance channel, through the existence of well 
defined PLF and
TLF like fragments, together with a quite large 
multiplicity of intermediate mass 
fragments, is observed \cite{bar02}. 
A total number of 2000 events is generated for each case and for the 
two iso-EOS considered.
In Figure \ref{mult} we report the 
IMF multiplicity distribution
for all reactions. We observe that the
SMF model is able to reproduce the general feature that a more
neutron rich combination enhances the IMF's multiplicity.  
 
To get a deeper insight into the nature of the fragmentation process,
we adopt an analysis of kinematical properties which was previously
employed in studies concerning dynamical fission or neck fragmentation
mechanisms \cite{stef1995,wilcz2005}.
 
\begin{figure}
\begin{center}
\includegraphics*[scale=0.4]{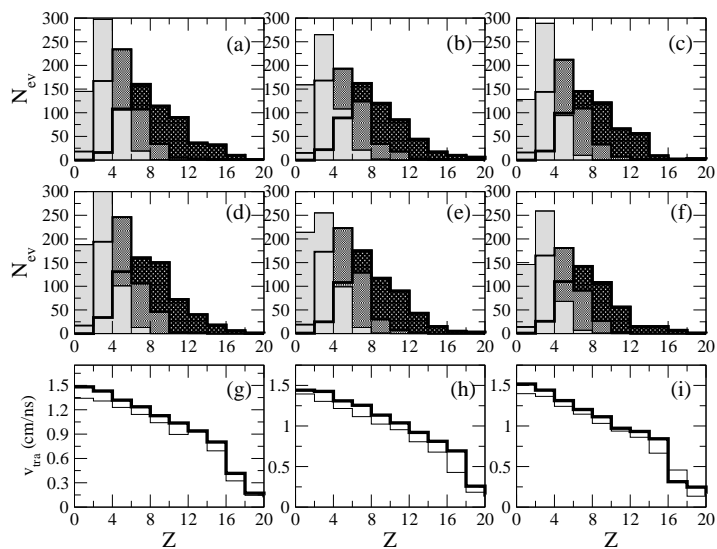}
\end{center}
\caption{Charge distribution of each IMF of the  hierarchy for Asysoft 
EoS (upper row) and
Asystiff EoS (middle row). HH combination: (a),(d),(g). 
EE combination: (b),(e),(h).
LL combination: (c),(f),(i).    
Average transverse velocity distribution as a function
of charge (bottom row) for Asysoft EoS (thick line) and Asystiff EoS 
(thin line).
All results refer to events with IMF multiplicity equal to three. 
The shading of 
the distributions
brightens according to the rank of the fragment in hierarchy.}
\label{dis5a}
\end{figure}

The asymptotic velocities of PLF and TLF-like 
residues define an intrinsic axis of the event by the vector 
${\bf{V}_r}= {\bf{V}}(H_1)-{\bf{V}}(H_2)$, always oriented from the 
second heaviest
fragment $H_2$ towards the heaviest one $H_1$.  
The intermediate mass fragments 
of each event are ordered in mass and the orthogonal and parallel components
of their asymptotic velocities, with respect to the intrinsic axis,
are determined. 

\begin{figure}
\begin{center}
\includegraphics*[scale=0.4]{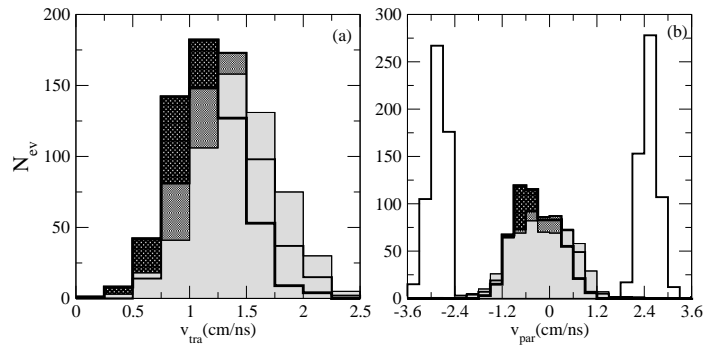}
\end{center}
\caption{(a): Transverse velocity, $v_{tra}$, distributions for 
fragmentation events with three IMF's.
(b): Parallel velocity distribution for fragmentation events with three IMF's.
Asysoft EoS and HH combination. The lightest shading corresponds to 
the lightest fragment
in the event.}
\label{vsoft}
\end{figure}

In Figure \ref{dis5a} 
the charge distributions corresponding to each order in hierarchy are 
shown
for the events
with three IMFs and all entrance channel combinations, HH, EE and LL 
respectively.
In all figures the histograms brighten as the rank of the IMF
increases. The heaviest IMF (the rank one in hierarchy) can have a charge up to
$Z=16-18$ and the distribution is centered around $Z=6-8$ while the 
lightest extends up to $Z=8$. 
The average transverse velocity in each charge bin was
calculated considering the contribution of all fragments, independently of 
the position
in hierarchy (see Figure \ref{dis5a} (g), (h) and (i)). 
One observes a steep decreasing trend
with the charge, in agreement with previous findings reported in 
\cite{liontiPLB625}.
Since similar results are obtained for all entrance channel combinations we
discuss in the following only the case of the HH system.
In Figure \ref{vsoft} 
we report the IMF transverse 
and parallel velocity distributions for Asysoft EoS.
We also plot the parallel velocity distributions of projectile and target like
residues. In agreement with our definition for the intrinsic axis of the event,
the velocity of the heaviest residue is always the positive one.
 
The transverse velocity distribution
shifts towards higher values with the position in the mass hierarchy, the
lightest fragment acquiring higher velocities. The same behavior is observed
for the class of events with four IMF's and also for Asystiff EoS.  
This hierarchy 
in the velocity perpendicular to the intrinsic axis emerges as a specific
signal characterizing the transition from multifragmentation to neck 
fragmentation.
It can be related to the peculiar geometrical configuration of the 
overlapping region and to its time evolution. 
Due to the Coulomb repulsion, neck fragments are essentially emitted on the
transverse plane. At variance light fragments, formed in a fast spinodal 
mechanism,  have shorter emission time and acquire
larger transverse velocity. It is interesting to notice that similar features
are observed in central reactions (see previous Section), but with respect
to the collective radial velocity.

\begin{figure}
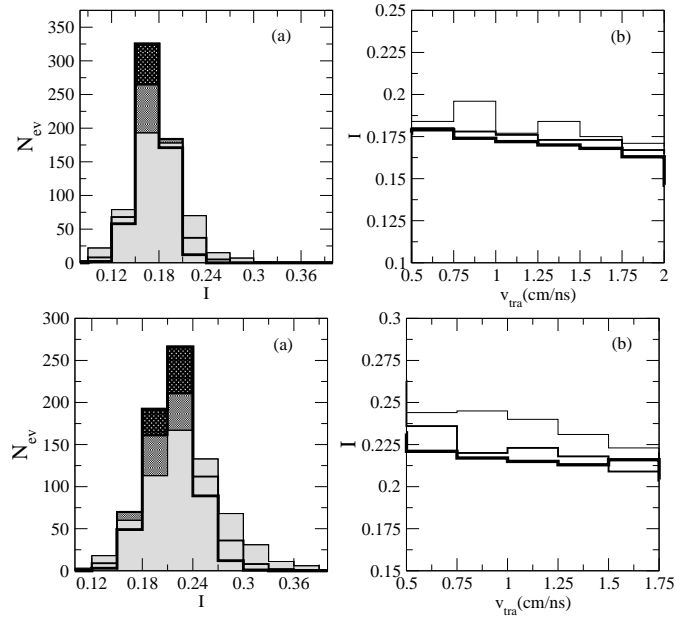

\centering
\includegraphics*[scale=0.4]{probe18a.eps}
\includegraphics*[scale=0.4]{probe18b.eps}
\caption{Asymmetry distribution of each fragment in hierarchy for fragmentation
events with three IMF's (a).  
Average asymmetry of each IMF in hierarchy as a function of transverse 
velocity 
for events with three IMF's (b).  
Upper Panel: Asysoft EoS and HH combination. Bottom Panel: Asystiff and HH
combination}
\label{iso13}
\end{figure}

The features discussed above are determined mainly by the isoscalar part of 
the equation of state, on top of which the symmetry energy may induce minor 
changes.
This explains the tiny difference between the two Iso-EoS. 
On the other hand, 
the properties related to the fragment isotopic content are 
directly influenced
by the symmetry energy term. We have extended our investigation
to this observable studying its dependence
on the IMF position in hierarchy, as well its correlation to transverse 
velocity, similarly to the analysis done for central collisions (see previous
Section). 
In Fig.\ref{iso13}(a), Upper Panel for Asysoft EoS and 
Bottom Panel for Asystiff EoS,
we report the asymmetry $I=(N-Z)/(N+Z)$ distribution of each IMF of the 
hierarchy.
The results refers to HH system whose initial asymmetry is $I=0.194$.
 
For Asysoft EoS the isospin distributions are centered at lower values
and have rather narrow widths, quite insensitive to the position
in hierarchy. At variance, for Asystiff EoS the centroids of 
the distributions
are closer to the initial value of the composite system and their 
broader widths
depend on the rank in the mass hierarchy. 
For both equations of state the lightest IMFs are more likely to acquire 
higher values of asymmetry pointing towards different production mechanisms or
formation conditions.

 We relate these features to the differences between the behavior of
the two Iso-EoS at subsaturation densities. Clearly, larger values of 
the symmetry energy (Asysoft case)
fasten the isospin distillation process and
all IMF's reach closer isospin values. 
On the other hand, 
for Asystiff EoS
fragments grow in low density, more 
charge asymmetric domains,
as a result of isospin migration towards the neck. The 
differences inside 
the hierarchy in this case point towards different formation time scales 
with the
lightest IMF finding a neutron rich environment, being the distillation 
process not 
efficient enough to lower the isotopic content  of all IMF's in the event. 
If they have lower
transverse velocity, remaining longer in a neutron-richer region, they 
will carry 
higher asymmetry.
The average fragment asymmetry, as a function of the transverse velocity, 
is shown in 
Fig.\ref{iso13} $(b)$, Upper Asysoft and Bottom 
Asystiff EoS.
A decreasing trend is manifest for all IMF's, more pronounced for 
the latter choice. In this case, 
for a given transverse velocity bin 
the asymmetry always increases with the rank in hierarchy.

These observations open new opportunities also from the 
experimental point of view.
An analysis of isospin dependent observables in correlation to the position
in mass hierarchy 
or kinematic observables 
can add new constrains on the behavior of
the symmetry energy below normal density and provide indications
about the formation of IMFs in the  low density 
and heated overlapping region at these impact parameters. 
 Recent experimental results reported by the CHIMERA
collaboration for the system $Sn+Ni$ at a lower energy ($35 AMeV$) 
 \cite{defilippo2010} suggest
the existence of the hierarchy in transverse velocity, as discussed here.
In these experiments it was also found that the lightest fragments are more 
asymmetric.
This sensitivity to the rank in the hierarchy can  then be related to a 
asystiff-like
behavior of the symmetry energy at sub-saturation densities.

\section{Isospin Effects at High Baryon Density: Effective Mass Splitting and 
Collective Flows}

At higher energies, where suprasaturation densities will be probed, all the 
observables under investigation are more affected by the Momentum Dependence
(MD) of the mean fields. This represents a further obstacle for 
the determination 
of $E_{sym}$ at high $\rho$, even because the MD has also an isoscalar 
contribution that is influencing the reaction dynamics, as shown in the 
previous Section.
 However, the possibility to have
access experimentally to several observables as a function of momentum in 
a wide range offers
the real possibility to disentangle density and momentum dependence of the 
symmetry potentials ($n/p$ effective masses).
Generally, all Relativistic Mean Field (RMFT) approaches give 
$m^{*}_{n}<m^{*}_{p}$, 
while in non-relativistic models generally $m^{*}_{n}>m^{*}_{p}$ with some 
exceptions \cite{ditoroAIP05,rizzo04}. Non-relativistic  
Brueckner-Hartree-Fock (BHF) \cite{bomblomb91} would indicate 
the last choice as 
the correct one, which means that the Lane Potential 
$U_{Lane} \equiv (U_p-U_n)/2\beta$, with $\beta$ asymmetry parameter, 
decreases 
with $k$.
In RMFT one finds the opposite trend but it lacks the effect of the finite 
range of the interaction. In Dirac-BHF both relativistic fields and the 
finite range effect are included, but there exists an ambiguity from the 
method used to project on the different Lorentz amplitudes\cite{fuchswci}.
One could judge that the microscopic approaches favour the relation 
$m^{*}_{n}>m^{*}_{p}$, but
in any case it is mandatory to show that this is the case in HIC
by means of a comparison with the available and forthcoming data.

Our Stochastic Mean Field ($SMF$)
 transport code has been implemented with 
 $Iso-MD$  symmetry potentials,
with a different $(n,p)$ momentum dependence, as discussed in detail
in Sect.2. This will allow to follow the dynamical
effect of opposite n/p effective mass splitting while keeping the
same density dependence of the symmetry energy   \cite{rizzo08,giordano10}.

We present here some results for  $^{197}Au+^{197}Au$
 reactions at $400AMeV$ \cite{giordano10}. For central 
collisions in the interacting zone we can reach baryon densities about
$1.7-1.8 \rho_0$ in a transient time of the order of 15-20 fm/c. The system 
is quickly expanding and the Freeze-Out time is around 50fm/c. At this time
we have a dominant Coulomb interaction among the reaction products. All the 
results presented here refer to this time step. Secondary decays of 
excited primary
fragments are not accounted for. In fact this will not affect too much the 
properties of nucleons and light ions at high transverse momenta 
mostly discussed 
in this work.

\begin{figure}
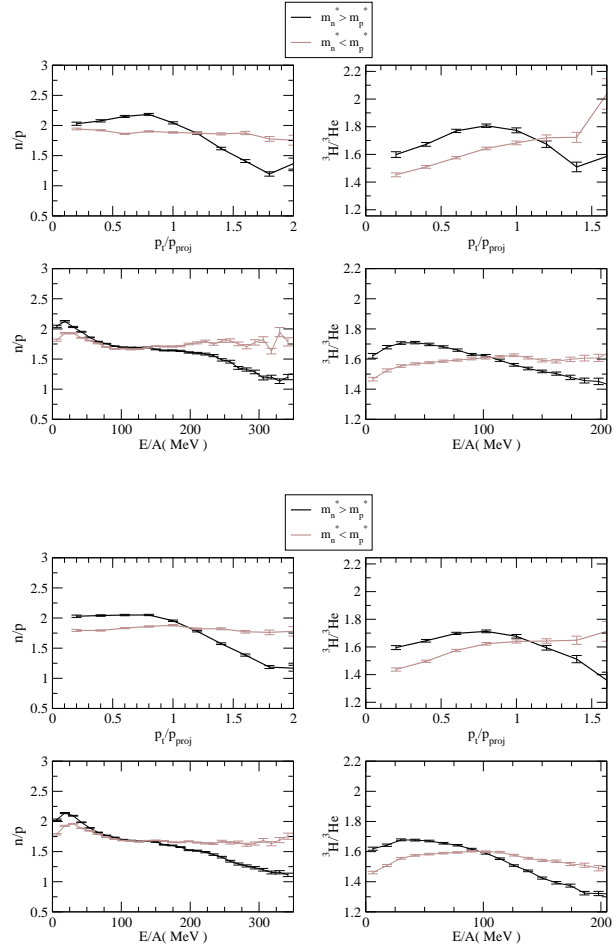

\centering
\includegraphics[width=8.0cm]{probe19a.eps}
\vskip 0.5cm
\includegraphics[width=8.0cm]{probe19b.eps}
\caption{197Au+197Au at 400AMeV, central collision.Isospin content of 
nucleon (left) and light ion (right) emissions vs. $p_t$ 
at midrapidity, $\mid y_0 \mid <0.3$, (upper) and
 kinetic energy (lower), for all rapidities,
for the two nucleon mass splitting choices.
Top Panels:
Asysoft; Bottom Panels: Asystiff.}
\label{ratios400}
\end{figure}

\subsection{Isospin Ratios of Fast Emitted Particles}

In Fig.\ref{ratios400}  we plot the  $(n/p)$ and
 $^3H/^3He$ yield 
ratios at freeze-out,
for two choices of Asy-stiffness and Mass-splitting, vs. transverse
momentum in a mid-rapidity selection (upper curves) and kinetic energy (
all rapidities, lower curves). In this way we can
separate particle emissions from sources at different densities, as discussed
in Sect 2.

We clearly observe the opposite effect of the 
different mass 
splitting in the low and high
momentum regions, as expected from Fig.\ref{mdpot}. 
E.g. in the $m_{n}^{*}<m_{p}^{*}$ case the neutrons see 
a less
repulsive potential at low momenta and a more repulsive one at high $p_t$. 
The curves
in the opposite mass-splitting show exactly the opposite behavior.
We note some interesting features: 

\noindent
i) The effect is almost not dependent
on the stiffness of the symmetry term. At high $p_t$, where particles
mostly come from high density regions, the larger repulsion seen by
neutrons in the Asy-stiff case, leading to an enhanced emission, is 
compensated by the larger Coulomb repulsion in the remaining matter, 
favoring proton emission. On the other hand, at low $p_t$, the sensitivity
to the Asy-stiffness is lost due to the mixing of sources at different 
densities, also for central rapidities, during the radial expansion. 

 \noindent
ii) The curves are crossing at
$p_t \simeq p_{projectile}= 2.13 fm^{-1}$. The crossing nicely corresponds
to the Fermi momentum of a source at baryon density $\rho \simeq 1.6 \rho_0$, 
 \cite{ditoroAIP05,rizzoPRC72,rizzo08}. 

\begin{figure}[t]
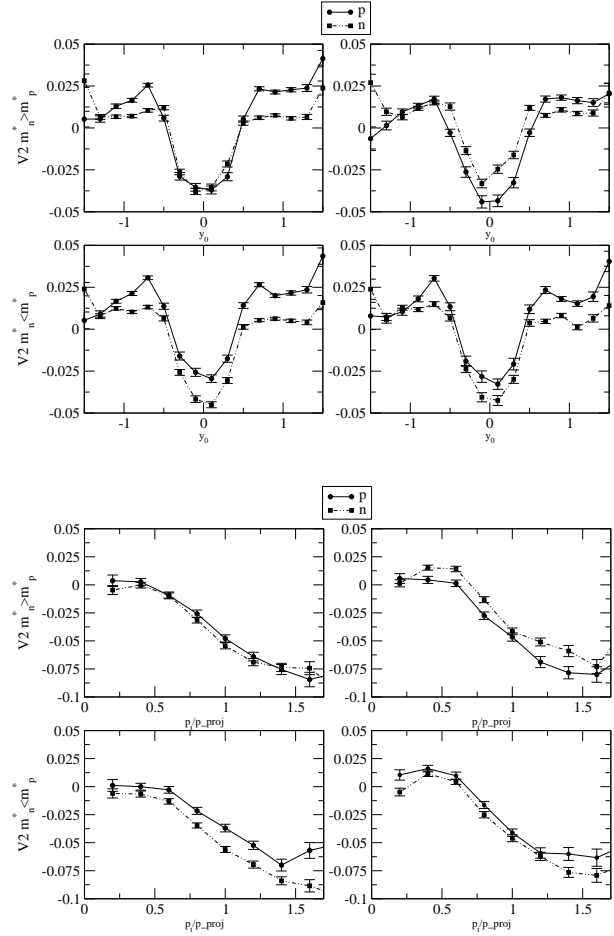

\centering
\includegraphics[width=8.0cm]{probe20a.eps}
\vskip 0.5cm
\includegraphics[width=8.0cm]{probe20b.eps}
\caption{
Proton (thick) and neutron (thin) $V_2$ flows in a 
semi-central
reaction Au+Au at 400AMeV.
Top Panels:
Rapidity dependence. 
Bottom Panels:
Transverse momentum dependence at midrapidity, $\mid y_0 \mid <0.3$.
Upper curves for $m_n^*>m_p^*$, lower curves for
the opposite splitting $m_n^*<m_p^*$. Left: Asystiff. Right: 
Asysoft.}
\label{v2ypt400}
\end{figure}

\begin{figure}[t]
\centering
\includegraphics[width=8.0cm]{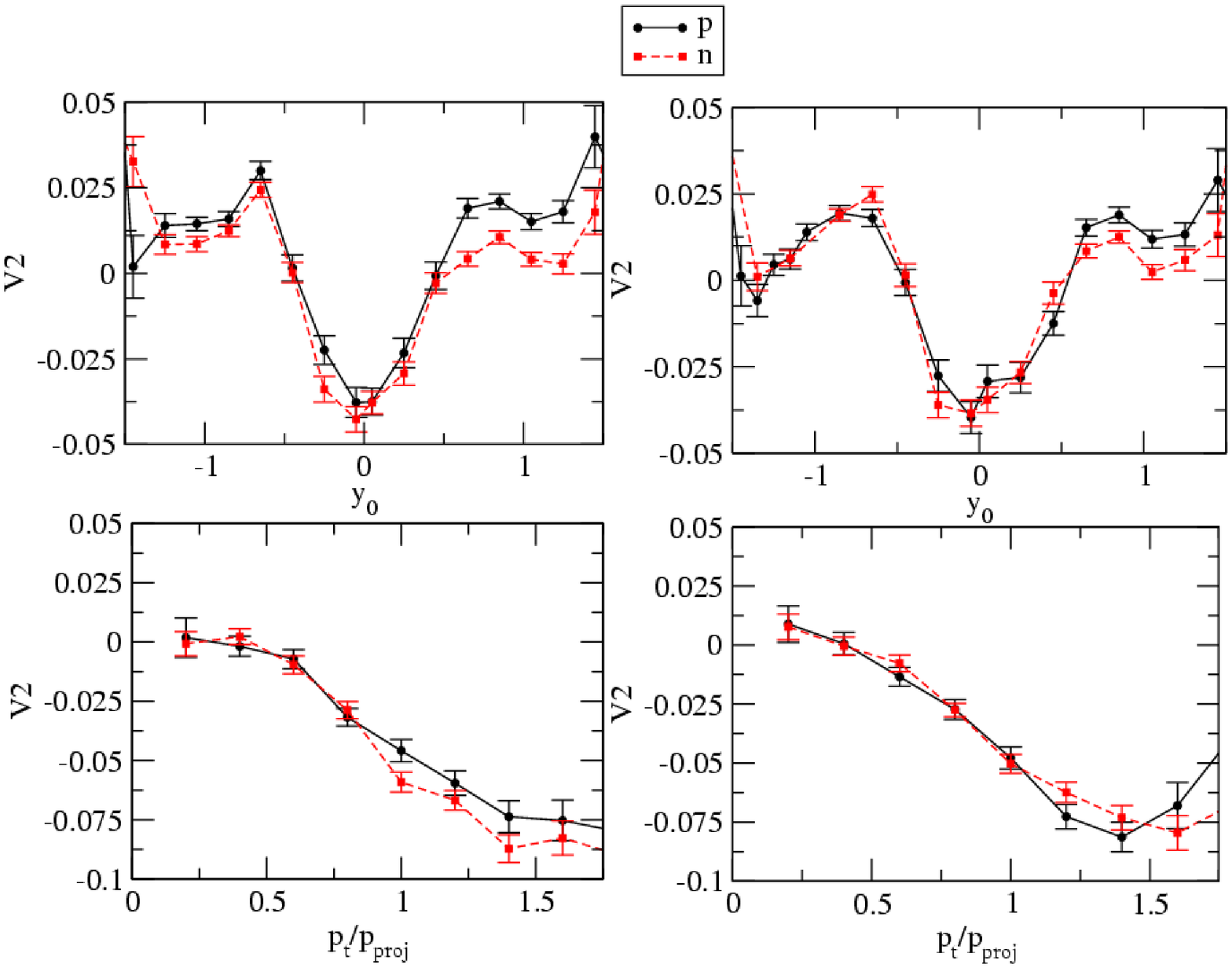}
\caption{
Proton (thick) and neutron (thin) $V_2$ flows in a 
semi-central
reaction Au+Au at 400AMeV, for equal (n.p)-effective masses.
Top Panels:
Rapidity dependence. 
Bottom Panels:
Transverse momentum dependence at midrapidity.
Left: Asystiff. Right: 
Asysoft.}
\label{nosplit}
\end{figure}

We remark that all the effects discussed before should be also present for the
$^3H/^3He$ yield ratios, more easily detected. Particularly interesting is the
predicted large increase at high $p_t$ in the $m_{n}^{*}<m_{p}^{*}$ choice. 
Some preliminary FOPI results seem to indicate this trend \cite{reis09}, 
but more data
are needed. It is encouraging that we already see a good $Iso-MD$ 
dependence of rather inclusive nucleon/cluster emission data. In presence of a 
good statistics for the detection of high $p_t$ particles, a further selection
at high azimuthal angles and central rapidities would certainly enhance the
sensitivity to the momentum dependence of the Symmetry Potentials. This
will introduce the discussion of isospin elliptic flows of the following
Subsection.

\subsection{Isospin Flows}

Isospin effects on collective flows have been studied within the $UrQMD$ 
transport model
in order to probe the influence of the symmetry repulsion at high densities
 \cite{qli0506,qli06}, here we focus the attention on the mass-splitting 
contributions.
For the same Au+Au reactions, in a semicentral selection, we present in Fig.
\ref{v2ypt400} (Top Panels) the rapidity dependence of 
$(n/p)$ elliptic flow $V_2$ for 
different choices of the Asy-stiffness and effective mass splitting. We 
observe the relevance of the latter: at mid-rapidity the neutron 
squeeze-out is much 
larger in the 
$m^*_n < m^*_p$ case independently of the stiffness of the symmetry term. 
We note however that in the Asysoft case we see an inversion of the 
neutron/proton squeeze-out
at mid-rapidity for the two effective mass-splittings. Good data 
seem to be suitable
to disentangle $Iso-MD$ potentials.

The mass-splitting effect is large at high $p_t$ (Bottom Panels), again in a 
mid-rapidity selection, 
as expected for particle emitted from higher density regions.
Here the results are also slightly depending on the Asy-stiffness,
 with large neutron $squeeze-out$ effects in the Asystiff case.

In order to have a clear idea of the relevance of the (n,p) mass 
splitting on the 
fast nucleon emissions we present in Fig.\ref{nosplit} the neutron/proton 
elliptic flows
for semicentral $Au+Au$ collisions at $400AMeV$ evaluated 
with the parametrizations giving $m^*_n=m^*_p$ 
for the $^{197}Au$ asymmetry $\beta \simeq 0.2$.
Now the isospin 
effects are only related to the different stiffness of the symmetry 
term at suprasaturation
density. We see that, at variance with the mass-splitting results of 
Fig.\ref{v2ypt400}, the
rapidity distributions (top panels) are not much affected, with a 
slightly larger neutron 
squeeze-out
in the Asystiff case. Consistently we see some difference in the 
transverse momentum dependence
at mid-rapidity (bottom panels) only at very large $p_t$.

Due to the difficulties in
measuring neutrons, we have analyzed the isospin sensitivity  
of light isobar flows, like $^3H$ vs. $^3He$ and so on.
We still see effective mass splitting effects, although 
slightly reduced. As in the nucleon elliptic flow, at mid-rapidity the 
triton squeeze-out is 
larger in the 
$m^*_n < m^*_p$ case independently of the stiffness of the symmetry term. 
Again in the Asysoft case we see an inversion of the  $^3H$ vs. $^3He$  
squeeze-out
at mid-rapidity for the two choices of the mass-splitting. 
Some larger mass-splitting effects can be seen at high $p_t$.
This should be well observed in the flow difference, with a good statistics.

We would expect the $Iso-MD$ effects to increase with beam energy, due to
the larger momenta of the emitted particles and to the reached higher
densities in the compression stage.
In this respect an interesting positive result is coming from
preliminary FOPI data on $^3H$ vs. $^3He$ flows, for Au-Au collisions
at beam energies
extended up to 1.5AGeV. The triton $V_2$ shows 
a larger squeeze-out
at mid-rapidity (in a relatively high transverse momentum selection)
 \cite{reis09}.
 New more accurate and exclusive 
data will soon appear. We
like to mention the new measurements that will be performed at SIS-GSI 
by the ASYEOS 
Collaboration \cite{lemmon09} and the new experiments planned at 
RIKEN-Tokyo and CSR-Lanzhou
also with unstable, more neutron-rich, beams.

\section{The Relativistic Structure of $E_{sym}$ }

The determination of $E_{sym}$ at high density $\rho> 1.5 \rho_0$ needs to 
exploit HIC's at beam energies
of $E/A\geq 400 MeV$ in order that the maximum density
is sufficiently high. For the theoretical approaches this shifts the 
energy regime of interest to a range where relativistic
effects can start to become significant. Hence relativistic, covariant 
approaches should be preferred;
nonetheless relativistic extensions of non-relativistic approaches represent 
a viable way even if they may mix relativity and many-body 
effects\cite{greco03}.
In this section we briefly recall the relativistic structure of $E_{sym}$
in effective hadronic models based on nucleons
interacting through mesonic fields. At the level presented here the 
structure of $E_{sym}$ 
is shared by the different models based on effective Lagrangians of 
Quantum-Hadro-Dynamics (QHD) \cite{qhd}, as the Relativistic Mean Field 
Theory (RMFT)
\cite{baranPR,liubo02,grecoPRC64}, the Density Dependent Relativistic Hartree 
approach (DDRH)\cite{fuchsPRC52}, the Effective Field Theory (EFT) 
density functional\cite{rusnak1997} as well as the microscopic 
Dirac-Brueckner-Hartree-Fock (DBHF)
\cite{dejong1998,alonso03}.
Therefore the following discussion allows to set a common language that 
will be useful
in the discussion about the meson production and the transition to quark matter
of the next two Sections.

In QHD the effect of nuclear interactions results
in in-medium modifications of the scalar effective masses and 
the energy-momentum
four-vector
\begin{equation}
\label{eff-mass}
m^{*}_{n,p}=m+\Sigma_{\sigma}\pm\Sigma_{\delta} \, \: , 
\: \: k^{*\mu}_{n,p}=k^{\mu}+ \Sigma^{\mu}_{\omega}\pm\Sigma^{\mu}_{\rho} \: ,
\end{equation}
where the different self-energies $\Sigma_i$ are labelled by the mesons 
representative of the
specific spin-isospin quantum numbers of field
(isoscalar, Lorentz scalar/vector, $\sigma/\omega$ and isovector,
 scalar/vector, $\delta/\rho$ fields). The upper and lower signs 
refer to neutrons and protons, respectively. In neutron-rich
asymmetric matter we systematically derive a $m_n^* < m_p^*$ condition,
 even in the non-relativistic limit \cite{baranPR,ditoroAIP05}.
 The isovector self-energies 
can be written as
\begin{equation}
\label{iso-selfene}
	\Sigma_{\delta}(n,p)= - f_{\delta} \rho_{s3} \, \: \:\:\: , 
\: \: \: \:\,
	\Sigma^{\mu}_{\rho}(n,p)= f_{\rho} j^{\mu}_{3} \: ,
\end{equation}
where $\rho_{s3}=\rho_{sn}-\rho_{sp}$ and 
$j^{\mu}_{3}=j^{\mu}_{n}-j^{\mu}_{p}$ are
the scalar isospin density and the isospin current, respectively. In the most
simple models of RMFT the coupling vertices $f_{i}$ are constant, 
 while in 
more sophisticated models. as e.g. the DDHF, they are density and 
even momentum dependent \cite{typel0305}.

\begin{figure}[th]
\centering
\includegraphics[width=5.5cm]{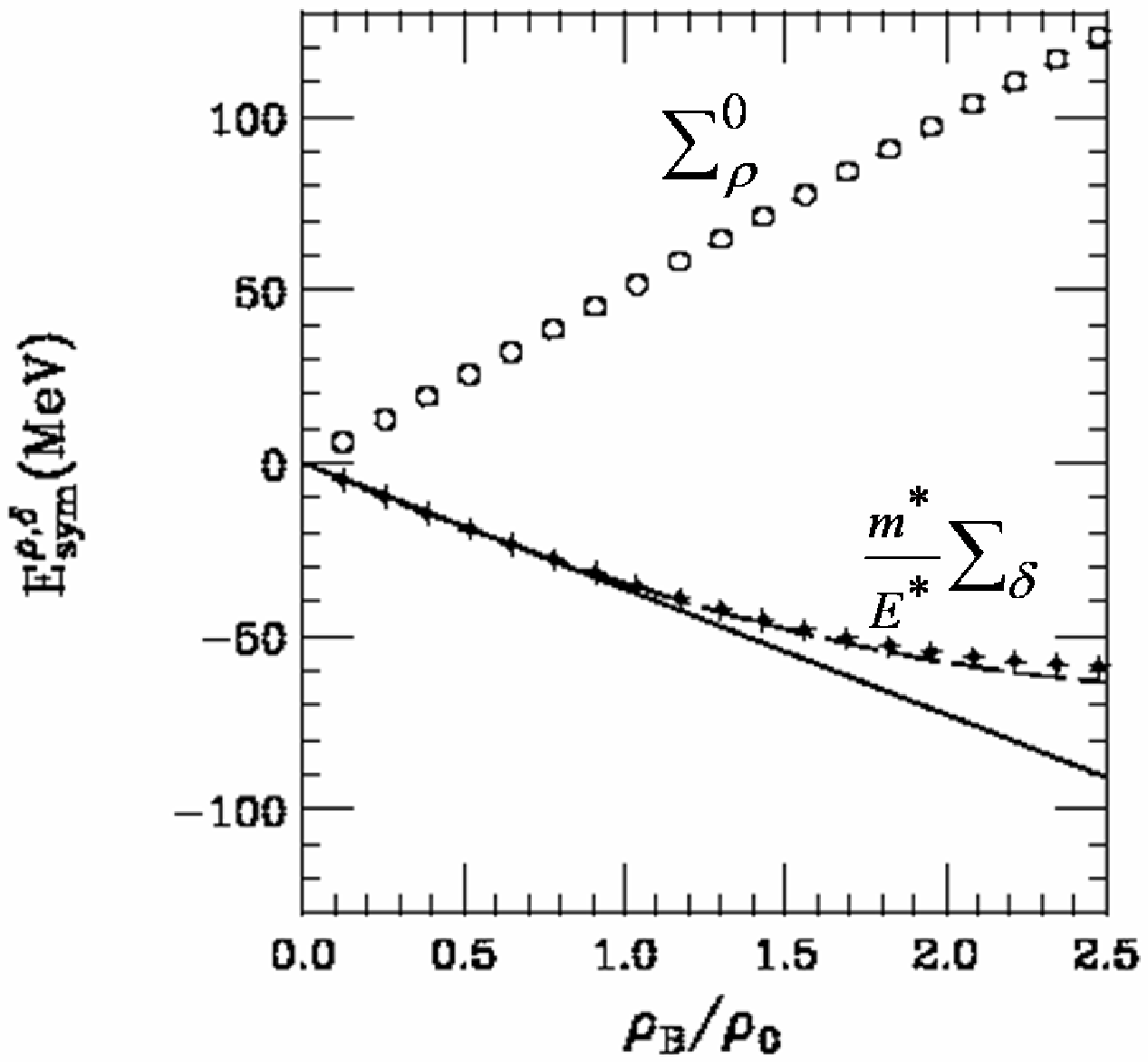}
\hspace*{10pt}
\includegraphics[width=5.5cm]{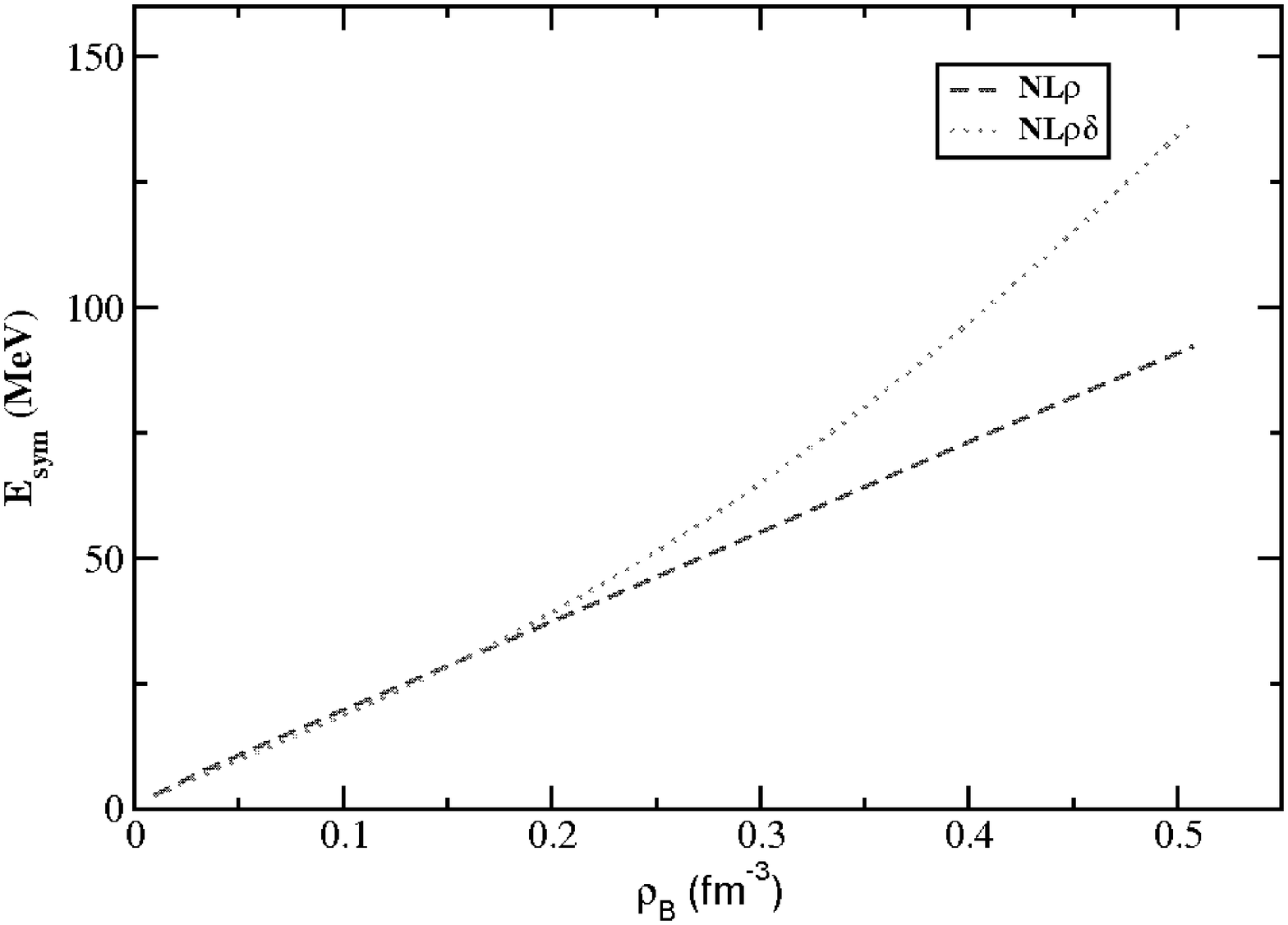}
\caption{
Left Panel: density dependence of 
scalar and vector isovector
self-energies in NL-RMFT; the solid line is to highlight the deviation from 
linearity. Right Panel: corresponding symmetry term with and without
the $\delta$ contribution.
}
\label{esymrel}
\end{figure}

The symmetry energy arising from such a structure can be written as
\cite{baranPR,liubo02}
\begin{equation}
\label{relesym}
E_{sym} = \frac{1}{6} \frac{k_{F}^{2}}{E_{F}^*} +
\frac{1}{2} \left[ f_{\rho} - f_{\delta}\left( \frac{m^{*}}{E_F^*} 
\right)^{2} \right] \rho_{B}~~
\simeq~~\frac{1}{6} \frac{k_{F}^{2}}{E_{F}^*} +
\Sigma^{0}_{\rho} +  \frac{m^{*}}{E_F^*} \Sigma_{\delta}
\end{equation}
with $E^* \equiv \sqrt{k^2 + m^*}$, 
where $f_\rho, f_\delta$ are the couplings of the nucleons to the effective 
isovector scalar and vector field. However
already from the Hartree-Fock theory as well as from the more complete 
DBHF approach it is clear that each spin-isospin channel receives 
contributions from
all the meson-like fields and the meson label is only indicative of the 
channel \cite{grecoPRC63}.
Eq.(\ref{relesym}) can be strictly derived only in RMFT,
but for our purpose the key point is that $E_{sym}$ arises
from a competition of two fields, one scalar and one vector, that 
at $\rho \sim 2\rho_0$ are of the order
of $\sim 100$ MeV, as is shown in
Fig.\ref{esymrel} (Left panel)  for a standard non-linear 
(NL)-RMFT model \cite{nonlinear}. 
Since the 
attractive $\delta$-contribution is proportional to the scalar density, 
reduced by a factor $m^*/E_F^*$ respect to the baryon 
density, the net effect will be a faster increase of the potential part 
of the symmetry energy at high densities \cite{baranPR,liubo02}, as shown
in the right panel of Fig.\ref{esymrel}.
This behaviour is shared by all relativistic approaches, with the difference 
that for example in DDRH the coupling $f_{\rho,\delta}$ depends on density, 
similarly in DBHF \cite{couplings}.

We stress the distinction between a scalar and a vector field because 
their balance in equilibrium conditions 
is generally dynamically broken for an evolving system. 
In Ref.\cite{greco03}
it has been shown that such an effect simulates a stiffer symmetry energy 
in the build-up of
collective flows. In fact, the difference of the force acting on a neutron 
and a proton moving with momentum
$\vec p$ can be written, after some approximations, as
\begin{equation}
\label{esym-hic}	
\frac{d\vec p_p}{d\tau}-\frac{d\vec p_n}{d\tau}\simeq 2\left[\gamma f_\rho 
-\frac{f_\delta}{\gamma} \right]\vec \nabla \rho_3 > \frac{4}
{\rho_B} E^{pot}_{sym} \vec \nabla \rho_3 \: ,
\end{equation}
where $\gamma$ is the Lorentz factor $E^*/m^*$. We can see that dynamically 
the vector field is enhanced
and the scalar field suppressed resulting in an effective $E_{sym}^*$ larger 
than $E^{pot}_{sym}$
as can be seen by comparing Eqs.(\ref{relesym}) and (\ref{esym-hic}).
Hence, the relativistic structure implies a modified relation between 
$E_{sym}$ and observables in HIC's.
Of course, one can simulate this relativistic structure with the help 
of a specific
momentum dependence, but this entangles many-body effects with 
relativistic ones.
A first analysis has shown that such effects are significant for $E/A > 1$ GeV
\cite{greco03}.

\section{Isospin Effects on Meson Production in Relativistic Heavy Ion
Collisions}
The phenomenology of isospin effects on heavy ion reactions
at intermediate energies (few $AGeV$ range) is extremely rich and 
the meson production can allow
a ``direct'' study of the Lorentz structure of the isovector interaction
in a high density hadron medium. We work within a relativistic transport frame,
beyond the cascade picture, 
 consistently derived from effective Lagrangians, where isospin effects
are accounted for in the mean field and collision terms.
We show that rather sensitive observables are provided by the 
pion/kaon production ($\pi^-/\pi^+$, 
$K^0/K^+$ yields). Relevant non-equilibrium effects
are stressed.

In this energy range lower mass mesonic states,
namely pions and kaons, can be excited. 
Therefore symmetry energy effects are transferred 
to the production of mesons.
In Ref.\cite{baoPRC67} it was suggested that the $\pi^-/\pi^+$ ratio 
is sensitive to $E_{sym}(\rho)$. It was claimed that the main effect 
consists in the
emission of nucleons during the evolution depending on $E_{sym}$ leading to 
a different $n/p$ content of the residual system, which in turn influences 
the formation of the different charge states of the pion.

\begin{figure}[th]
\centering
\includegraphics[height=2.3in,width=4.0in]{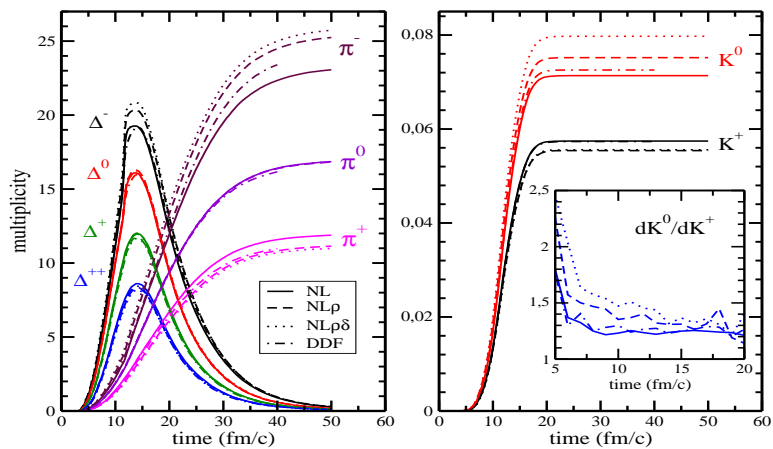}
\caption{Time evolution of the $\Delta^{\pm,0,++}$ resonances and pions 
$\pi^{\pm,0}$ (left),
and  of kaons ($K^{+,0}$(right) for a central ($b=0$ fm impact parameter)
Au+Au collision at 1 AGeV incident energy. Transport calculation using the
$NL, NL\rho, NL\rho\delta$ and $DDF$ models for the iso-vector part of the
nuclear $EoS$ are shown. The inset shows the differential $K^0/K^+$  ratio
as a function of the kaon emission time.}
\label{mesontime}
\end{figure}

Even if this mechanism seems quite straightforward one may doubt its 
real effectiveness. Indeed, the idea of
using pions to determine the symmetric part of the EoS has suffered 
from the fact that pions
strongly interact with nucleons and are produced during the whole 
evolution of the collision system making it difficult to associate
their production to a specific density reached during the collision.
In that context it was suggested by Aichelin and Ko \cite{aichko85} 
that kaons are a better probe of the EoS. The reason is twofold.
 Kaons have a higher threshold energy, hence they are produced only in the
high density phase. Moreover, once produced they interact weakly with 
nucleons and their width with respect
to the mass is quite small making a quasi-particle approximation more reliable.
After nearly 20 years the effort to determine the symmetric EoS by kaon 
production has been successful and is summarized
in Ref.\cite{fuchs06}.
Following the same line of thinking the Catania group has suggested to 
investigate the $K^0/K^+$ ratio
 as a better probe of the $E_{sym}$ at high density 
\cite{ferini05,ferini06,erice08}
(the other isospin pair with anti-kaons $\bar{K}_0/\bar{K}^-$ suffers 
from the strong
coupling to the medium).

Using a relativistic transport approach 
(Relativistic Boltzmann-\"Uhling-Uhlenbeck, RBUU
\cite{FuchsNPA589}) we have analyzed
pion and kaon production in central $^{197}Au+^{197}Au$ collisions in
the $0.8-1.8~AGeV$ beam energy range, comparing models with the same 
``soft'' EoS for symmetric
matter and with different effective field choices for $E_{sym}$.
Fig.\ref{mesontime} reports  the temporal evolution of $\Delta^{\pm,0,++}$
resonances, pions ($\pi^{\pm,0}$) and kaons ($K^{+,0}$)
for central Au+Au collisions at $1AGeV$\cite{ferini06,Pra07,wolter_erice08}
It is clear that, while the pion yield freezes out at times of the order of
$50 fm/c$, i.e. at the final stage of the reaction (and at low densities),
kaon production occurs within the very early (compression) stage,
 and the yield saturates at around $15 fm/c$, when the nucleon
and $\Delta's$ densities reach their maximum value.
From Fig.\ref{mesontime} we see that the pion multiplicities are
moderately dependent on the isospin part of the nuclear mean field.
However, a slight increase (decrease) in the $\pi^{-}$ ($\pi^{+}$)
multiplicity is observed when going from a
NL-RMFT model without isovector fields (NL) to one  with a $\rho$ meson 
($NL\rho$) and with $\rho$ and $\delta$ mesons ($NL\rho\delta$).
 This trend is more pronounced for kaons, see the
right panel, due to the high density selection of the source and the
proximity to the production threshold. Moreover isospin effects enter twice 
in the two-step production of kaons, see \cite{ferini05,ferini06}.

\begin{figure}[th]
\centering
\includegraphics[height=2.3in,width=2.6in]{probe24a.eps}
\hspace*{10pt}
\includegraphics[height=2.3in,width=2.8in]{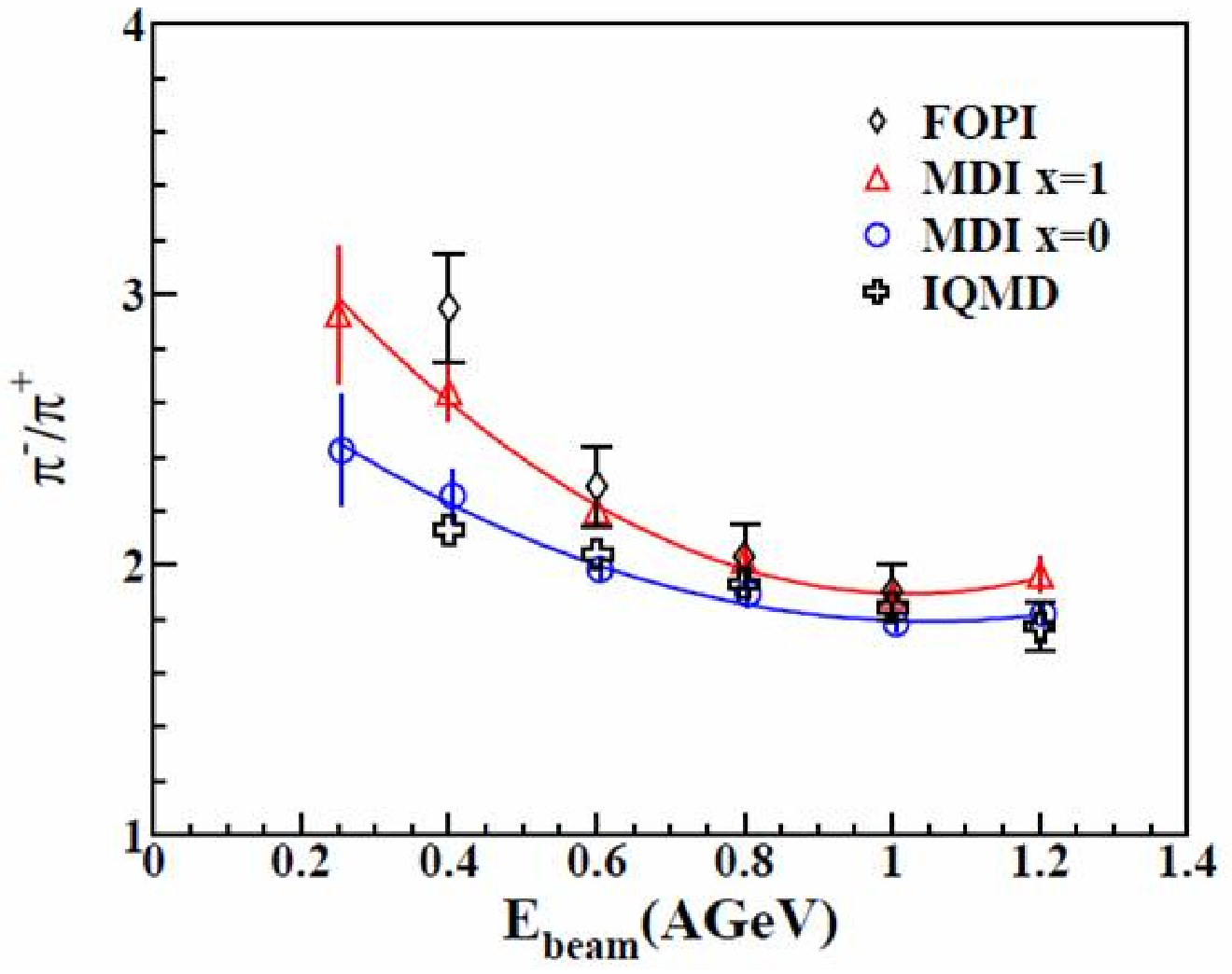}
\vspace*{1pt}
\caption{Left: Excitation function of the $\pi^-/\pi^+$ and $K^0/K^+$ ratios 
for $Au+Au$. RBUU results for different
behavior of $E_{sym}(\rho)$ \cite{ferini06}. 
Right: Results for $Au+Au$ in IBUU04 for a soft (x=1) 
and a stiff (x=0) $E_{sym}$ \cite{xiaosoft}
compared to the data from FOPI and a calculation with IQMD.
Taken from \cite{reisdorf07}.
}
\label{mesonbeamen}
\end{figure}

In Fig.\ref{mesonbeamen}(left) the pion and kaon ratios from these 
calculations are 
shown as a function of beam energy. The ratios decrease with beam energy, 
because the relative effects of mean fields and thresholds become less 
important. However, the greater sensitivitiy of the kaon ratios is seen 
clearly, since kaons emerge from high density regions (see the inset in 
the right panel of Fig.\ref{mesontime}. In this respect transverse momentum
selections of the $\pi^{-,+}$ yields would be very useful.

We have to note that in a previous study of kaon production in excited nuclear
matter the dependence of the $K^{0}/K^{+}$ yield ratio on the effective
isovector interaction appears to be much larger (see Fig.8 of 
ref.\cite{ferini05}).
The point is that in the non-equilibrium case of a heavy ion collision
the asymmetry of the source where kaons are produced is in fact reduced
by the $n \rightarrow p$ ``transformation'', due to the favored 
$nn \rightarrow p\Delta^-$ processes. This effect is almost absent at 
equilibrium due to the inverse transitions, see Fig.3 of 
ref.\cite{ferini05}. Moreover in infinite nuclear matter even the fast
neutron emission is not present. 
This result clearly shows that chemical equilibrium models can lead to
uncorrect results when used for transient states of an $open$ system.

At present there are essentially no data for kaons, while there are some 
for pions and $\pi^-/\pi^+$ ratios from
FOPI collaboration \cite{reisdorf07}. Within an isospin and momentum 
dependent transport model
it has been claimed that an agreement with data can be achieved only if 
a very soft $E_{sym}(\rho)$ is
employed \cite{xiaosoft}. Such a finding appears in strong disagreement 
with other studies
that exploit the elliptic flow to extract the slope of $E_{sym}$ around 
and above the saturation density \cite{trautPPNP09}. Moreover a recent
transport evaluation of the $\pi$ ratio, also in a non-relativistic frame, 
is reaching just
opposite conclusions \cite{zhao10}.

The effect described in Ref.\cite{xiaosoft} essentially is due to the fact 
that a stiff
$E_{sym}$ causes a neutron-rich emission of nucleons in the early stages 
of the reaction leaving the system
too symmetric in isospin content to reproduce the $\pi^-/\pi^+$ ratio of FOPI.
On the other hand, if one employs an $E_{sym}(\rho)$ that just 
above $\rho_0$ decreases with density generating an isospin
force that is attractive for the neutrons it is possible to get close 
to the data,
see Fig.\ref{mesonbeamen} (right).
However, we note that it is mandatory to check if one can reproduce 
at the same time the
$n/p$ emission as a function of $p_t$ especially at the high $p_t$ relevant 
for pion production, since this represents the complementary signal 
for the soft  $E_{sym}$, see Sect.5.
This would be a first test of the claim for a very soft $E_{sym}$ and
in case of failure it will
provide evidence that there are other mechanism determining the in-medium
meson production.
Indeed, we already know that there are at least other three effects competing
with the mean field effect on the $n/p$ emission.
These are the so-called "`threshold effect"'
emphasized by the calculation of Ref.s \cite{ferini05,ferini06,erice08},
the momentum dependence of the isospin-dependent cross sections, and the
 the isospin mass shift of pions due to the coupling to $\Delta-$hole 
excitation\cite{ko_pispectr10}. These will be discussed in the next
subsection..

We finally note that a decreasing symmetry term at high density, becoming 
even negative
 above $\simeq 3\rho_0$, implies more fundamental problems for the stability 
of a dense isospin asymmetric matter. We can have a collapse as well as
$n/p$ separation instabilities, since the isovector Landau parameter
can reach large negative values \cite{baranPR}. No evidence of such effects
is present in compact stars as well as in heavy ion collisions.

There are circumstantial reasons to be careful. Indeed, 
the physics
involved in the in-medium particle production has many aspects and 
a comprehensive and self-consistent
approach is necessary before extracting the isospin dependence of the
 interaction.

\subsection{Isospin dependence of thresholds and spectral functions}
\label{sec:threshold}

The "`threshold effect"' is due to the fact that the masses of nucleons and
Delta's are modified in the medium.
The unknown self-energies of the delta's are usually specified in terms 
of the neutron and proton ones by the use
of Clebsch-Gordon coefficients for the isospin coupling of the $\Delta's$ 
to nucleons
\cite{ferini05,baoPRC67}.
These medium modifications are isospin dependent, Eq.(\ref{iso-selfene}).
This influences the phase-space available for meson production in a 
nucleon-nucleon (NN) collision because it modifies the difference 
between the invariant energy
in the entrance channel $s_{in}$ with respect to the production threshold 
$s_{th}$.
This effect is, of course,
present in general for all meson productions, but for brevity we 
concentrate here on
one inelastic channel: $nn\rightarrow p\Delta^-$, mainly responsible 
for $\pi^-$ production.
From Eq.(\ref{eff-mass}) the invariant energy in the entrance channel 
and the threshold energy are given, respectively, by

\begin{eqnarray}
\label{threshold}
\sqrt{s_{in}}/2= \left[ E^{*}_{n}+\Sigma^{0}_{n}\right]\stackrel{p=0}
{\rightarrow}\left[m^{*}_{N}
+\Sigma^{0}_{\omega}+\Sigma^{0}_{\rho}+\Sigma_{\delta}\right]>m^{*}_{N}+
\Sigma^{0}_{\omega} \nonumber\\
\sqrt{s_{th}}= \left[m^{*}_{p}+m^{*}_{\Delta^-}+\Sigma_{0}(p)+\Sigma_{0}
(\Delta^-)\right]=
\left[m^{*}_{N}+m^{*}_{\Delta}+2 \Sigma_{\omega} \right]
\end{eqnarray}

where $m^{*}_{N}, m^{*}_{\Delta}$ are the isospin averaged values.
The last equality for the threshold energy is valid due to the prescription 
for the Delta self-energies, which leads to an exact compensation 
of the isospin-dependent parts, hence
the threshold $s_{th}$ is not modified by isospin dependent self-energies.
In general in a self-consistent many-body calculation higher order effects 
can destroy this exact balance.
But this is not so important for our qualitative discussion, since there 
will always be a compensating effect in $s_{th}$.
On the other hand, the energy available in the entrance channel, $s_{in}$, is
shifted in an explicitely isospin dependent way by the in-medium 
self-energy $\Sigma^{0}_{\rho}+\Sigma_{\delta}>0$.
Especially, the vector self-energy gives a positive
contribution to neutrons that increases the difference $s_{in}-s_{th}$ 
and hence increases the cross section of the
inelastic process due to the opening up of the phase-space, expecially 
close to threshold, since the intermediate $\Delta$ resonance will be 
better probed.

A similar modification but opposite in sign is present in $s_{in}-s_{th}$  
for the $pp\rightarrow n\Delta^{++}$
channel that therefore is suppressed by the isospin effect on the 
self-energies.
Hence, due to the described threshold effect the ratio $\pi^{-}/\pi^{+}$ 
increases with the stiffness
of $E_{sym}$ which is associated with a large 
$\Sigma^0_{\rho}+\Sigma_{\delta}$. This is at the
origin of the result in
Fig.\ref{mesonbeamen}(left). Of course, the RBUU calculation contains also
isospin contributions in the mean field but the final result 
appears to be dominated by the
threshold effect, in particular at lower energies.

However there are at least two other physical aspects that 
have to be considered.
One is the fact that the self-energies in the implementation of RBUU 
are not explicitly momentum
dependent as in DBHF and as the study of the optical potential shows.
Thus we return to the importance of the isospin momentum dependence 
discussed in
Sect.5.
Because we neglect the dependence of the self-energies on momentum,
the difference $s_{in}-s_{th}$ in Eq.(\ref{threshold}) increases strongly 
with the momentum
$p$. The problem is directly related to the strong energy dependence 
of the optical
potential in RMFT.
Therefore it is likely that a more realistic calculation
which includes a $\Sigma_{\rho,\delta}({\rho,p})$ will reduce the 
effect seen in
Fig.\ref{mesonbeamen}.
In addition a fully consistent treatment should include an optical potential
also for the pions.
It can be envisaged that a stronger sensitivity to $\Sigma({\rho,p})$ can 
be seen
looking at $\pi^{-}/\pi^{+}$ as a function of the pion transverse energy.
Furthermore in a fully self consistent many-body calculation the
$\Sigma$'s will affect also the isospin dependence of the cross sections
for inelastic processes.

Another effect that is certainly present is the modification
of the pion spectral function which in an asymmetric medium becomes 
isospin dependent.
This has been recently pointed out in Ref.\cite{ko_pispectr10}
using a pion interaction in the nuclear medium given
by chiral perturbation theory. 
The effect seems to go in the direction
of smaller $\pi^-/\pi^+$ ratios.
Therefore it would be important to include such pion in medium
interaction in the 
transport models,
even if it means to go beyond the simple quasi-particle approximation
 \cite{cassing09}.

\begin{figure}
\begin{center}
\includegraphics[scale=0.45]{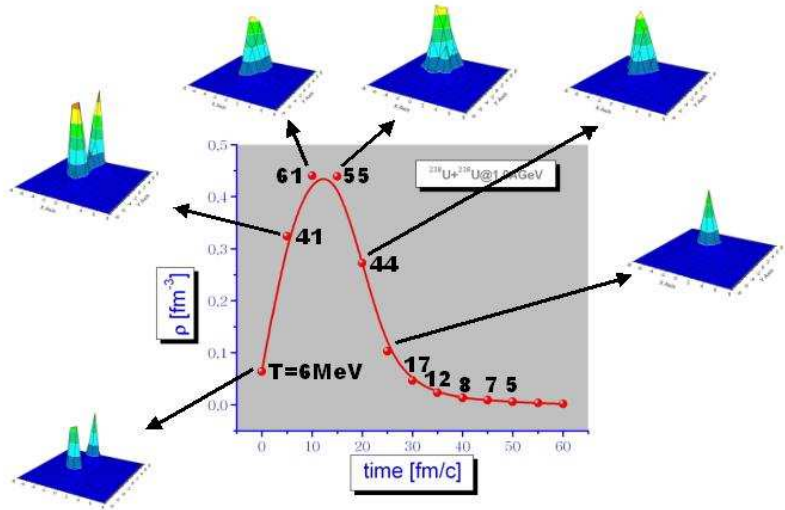}
\vskip -0.7cm
\caption{$^{238}U+^{238}U$, $1~AGeV$, semicentral. Correlation between 
density, 
temperature (black values), momentum
thermalization (3-D plots), inside a cubic cell, 2.5 $fm$ wide, in the center
of mass of the system.}
\label{figUU}
\vskip -0.7cm
\end{center}
\end{figure}

\section{Hadron-Quark Transition at High Baryon and Isospin Density}

In order to check which kind of matter we are probing in the $AGeV$ 
beam energy 
range, 
also having in mind the possibility of observing 
some precursor signals
of a new physics even in collisions of stable nuclei at
intermediate energies, we have performed some event simulations for the
collision of very heavy, neutron-rich, elements. We have chosen the
reaction $^{238}U+^{238}U$ (average proton fraction $Z/A=0.39$) at
$1~AGeV$ and semicentral impact parameter $b=7~fm$ just to increase
the neutron excess in the interacting region. 
In  Fig.\ref{figUU} we report the evolution of momentum distribution
and baryon density in a space cell located in the c.m. of the system,
 see \cite{ditorodec}.
We see that after about $10~fm/c$ a local
equilibration is achieved.  We have a unique Fermi distribution and
from a simple fit we can evaluate the local temperature 
(black numbers in MeV).
We note that a rather exotic nuclear matter is formed in a transient
time of the order of $10~fm/c$, with baryon density around $3-4\rho_0$,
temperature $50-60~MeV$, energy density $500~MeV~fm^{-3}$ and proton
fraction between $0.35$ and $0.40$, see \cite{ditorodec}. 
We can expect some chance of observing signatures of a transition from 
hadron to quark matter
at high baryon and isospin density and relatively low temperature,
as discussed in detail in this final Section.

\subsection{Isospin Effects on the Hadron-Quark Transition at High Density}

Several suggestions are already present about the possibility of 
interesting isospin effects on the transition to a mixed hadron-quark phase 
at high baryon density \cite{muller,ditorodec,erice08}. This 
seems to be a
very appealing physics program for the new facilities, FAIR at 
GSI-Darmstadt \cite{fair} and NICA at JINR-Dubna \cite{nica}, 
where heavy ion beams (even unstable, with large isospin asymmetry)
will be available with good intensities in the 1-30 AGeV energy region.

The weak point of those predictions is the lack of a reliable equation of 
state that can describe with the same confidence the two phases, hadronic and
deconfined. On the other hand this also represents a strong theory motivation
to work on more refined effective theories for a strong interacting matter.

\begin{figure}
\centering
\includegraphics[scale=0.40]{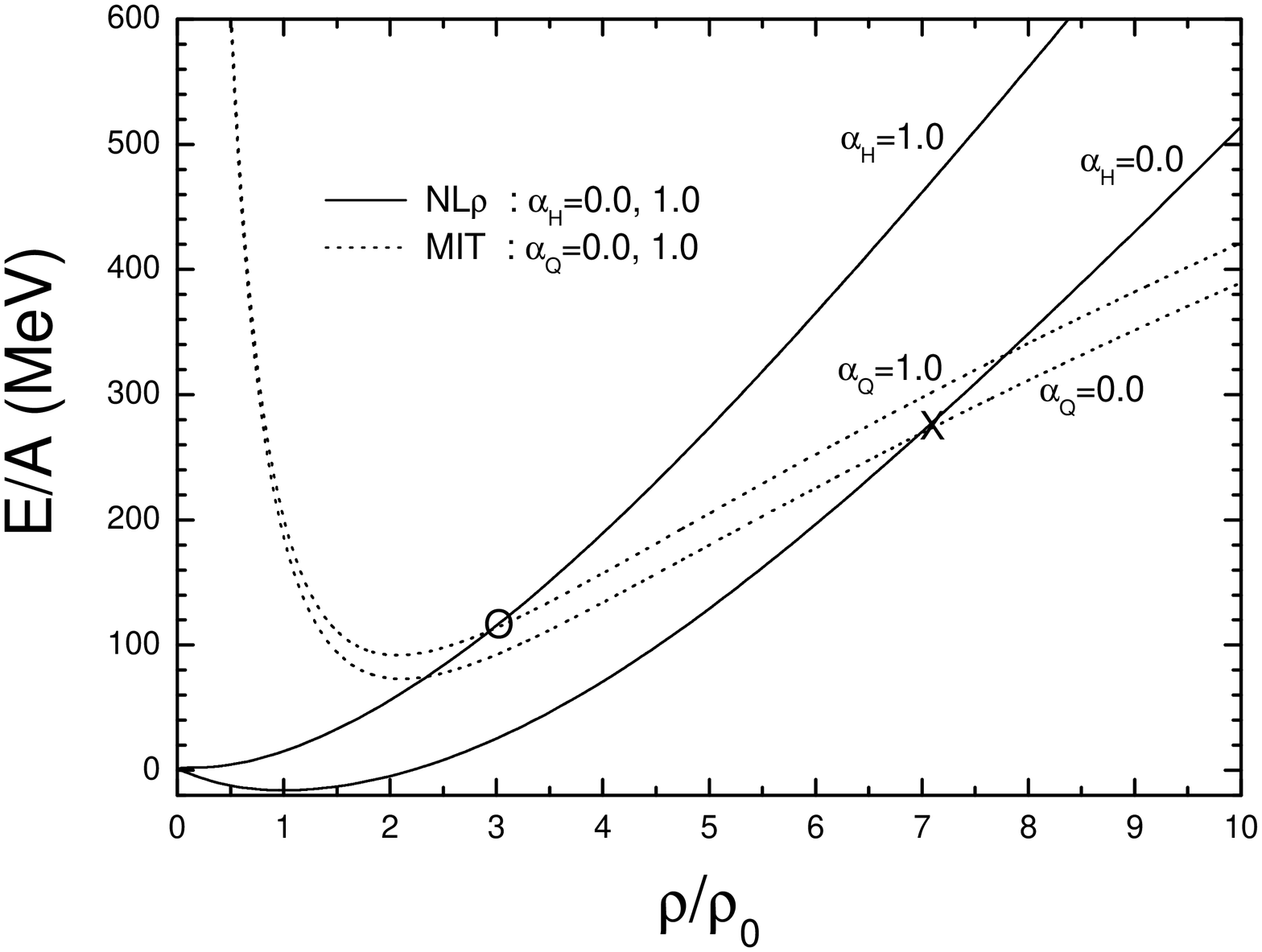} 
\caption{Zero temperature $EoS$ of Symmetric/Neutron Matter: 
Hadron ($NL \rho$), solid lines,
vs. Quark (MIT-Bag), dashed lines. $\alpha_{H,Q}$ represent the isospin 
asymmetry parameters respectively of the hadron,quark matter:
$\alpha_{H,Q}=0$, Symmetric Matter; $\alpha_{H,Q}=1$, Neutron Matter.
}
\label{isoparton} 
\end{figure}

A nice qualitative argument in favor of noticeable isospin effects on the 
hadron-quark 
transition at high density can be derived from
the Fig.\ref{isoparton}, where we compare typical Equations of State (EoS) for 
Hadron (Nucleon) and Quark 
Matter, at zero temperature, for symmetric 
($\alpha \equiv (\rho_n-\rho_p)/\rho_B \equiv -\rho_3/\rho_B=0.0$) 
and neutron matter 
($\alpha=1.0$), where $\rho_{n,p}$ are the neutron/proton densities and
$\rho_B=\rho_n+\rho_p$ the total baryon density.
For the Hadron part we adopt a Relativistic Mean Field EoS
(Sects.6-7) 
with non-linear 
terms and an effective $\rho-meson$ coupling for the isovector part, 
largely used 
to study isospin effects in relativistic $HIC$ (Sect.7) \cite{nlrho}.

The energy density and the pressure for the quark phase are given by 
the MIT Bag model \cite{MIT} (two-flavor case), with the bag constant
 taken as a rather 
standard value from the hadron spectra ($B=85.7~MeV~fm^ {-3}$, no density 
dependence) \cite{ditorojpg}.

The transition to the more repulsive quark matter will appear around the
crossing points of the two EoS. We see that such crossing for symmetric matter
($\alpha_H=\alpha_Q=0.0$) is located at rather high density, 
$\rho_B \simeq 7\rho_0$, while for pure neutron matter 
($\alpha_H=\alpha_Q=1.0$) it is moving down to $\rho_B \simeq 3\rho_0$.   
Of course Fig.\ref{isoparton} represents just a 
simple energetic argument to
support the hadron-quark transition to occur at lower baryon densities
for more isospin asymmetric matter. In the following we will
rigourously consider the case of a first order phase transition in the Gibbs
frame for a system with two conserved charges (baryon and isospin), in order 
to derive more detailed results. 
Since the first order phase transition presents a jump in the energy, we can 
expect the mixed phase to start at densities even before the crossing points
of the Fig.1. The lower boundary then can be predicted at relatively low 
baryon densities for asymmetric matter, likely
reached in relativistic heavy ion collisions and in compact stars.
In fact this
point is certainly of interest for the structure of the crust and the inner 
core of 
Neutron Stars (NS), e.g. see refs. \cite{burgio02,nicotra06} and the review 
\cite{page06}. We like to mention a very recent estimation of the NS
Mass-Radius correlation ($1.6M_{sun}~-~10Km$) with high confidence level,
that could indicate a transition to quark matter in the inner core
\cite{baym10}.

We finally note that the above conclusions are rather independent on the 
isoscalar 
part of the used Hadron
EoS at high density, that is chosen to be rather soft in agreement with 
collective flow and
kaon production data \cite{daniel02,fuchs06}.
 
In the used Bag Model no gluon interactions, the $\alpha_s$-strong coupling 
parameter, are 
included. We remark that this in fact would enhance the above effect, since it
represents an attractive correction for a fixed B-constant, 
see \cite{bmuller95}. A reduction of the Bag-constant with increasing 
baryon density, as suggested by various models, see ref.\cite{burgio02}, 
will also go in the direction of an ``earlier'' (lower density) transition,
as already seen in ref.\cite{ditorodec}.
At variance, the presence of
explicit isovector interactions in the quark phase could play an 
important role, 
as shown in the following also for other isospin properties inside the mixed 
phase.

\subsection{Isospin effects on the Mixed Phase}

We can study in detail the isospin dependence of the transition 
densities
\cite{muller,ditorodec,erice08}.

The structure of the mixed phase is obtained by
imposing the Gibbs conditions \cite{Landaustat} for
chemical potentials and pressure and by requiring
the conservation of the total baryon and isospin densities:

\noindent
\begin{eqnarray}\label{gibbs}
&&\mu_{B}^{H}(\rho_B^H,\rho_3^H,T)=\mu_{B}^{Q}(\rho_B^Q,\rho_3^Q,T)~, 
\nonumber \\
&&\mu_{3}^{H}(\rho_B^H,\rho_3^H,T)=\mu_{3}^{Q}(\rho_B^Q,\rho_3^Q,T)~, 
\nonumber \\
&&P^{H}(T)(\rho_B^H,\rho_3^H,T)=P^{Q}(T)(\rho_B^Q,\rho_3^Q,T)~, 
\nonumber \\
&&\rho_{B} = (1-\chi)\rho_B^H + \chi \rho_B^Q ~,
\nonumber \\
&&\rho_{3}= (1-\chi)\rho_3^H + \chi \rho_3^Q ~,
\end{eqnarray}

where $\chi$ is the fraction of quark matter in the mixed phase 
and T is the temperature.

The consistent definitions for the densitites and chemical potentials
in the two phases are given by :

\begin{eqnarray}\label{hadronrhomu}
&&\rho_{B}^{H}=\rho_{p}+\rho_{n}, ~~~~\rho_3^H=\rho_p-\rho_n~,
\nonumber \\
&&\mu_B^{H} = \frac{\mu_p + \mu_n}{2}, ~~~~~
\mu_3^{H} = \frac{\mu_p - \mu_n}{2}~,
\end{eqnarray}
for the Hadron Phase and

\begin{eqnarray}\label{quarkrhomu}
&&\rho_B^{Q}=\frac{1}{3}(\rho_{u}+\rho_{d})~,
~~~~~~ \rho_3^{Q}=\rho_{u}-\rho_{d}~,
\nonumber \\
&&\mu_B^{Q}=\frac{3}{2}(\mu_u + \mu_d),~~~~
\mu_3^{Q} =\frac{\mu_u - \mu_d}{2}~,
\end{eqnarray}

for the Quark Phase.

The related asymmetry parameters are:
\begin{equation}
\alpha^{H} \equiv -\frac{\rho_3^{H}}{\rho_B^{H}} =
\frac{\rho_n-\rho_p}{\rho_n+\rho_p}~,~~~~~~
\alpha^{Q} \equiv -\frac{\rho_3^{Q}}{\rho_B^{Q}} =
 3 \frac{\rho_d-\rho_u}{\rho_d+\rho_u}~.
\end{equation}

Nucleon and quark chemical potentials, as well as the pressures in the 
two phases, are directly derived from the respective EoS.

In this way we get the $binodal$ surface which gives the phase coexistence 
region
in the $(T,\rho_B,\rho_3)$ space. 
For a fixed value of the
total asymmetry  $\alpha_T=-\rho_3/\rho_B$ 
 we will study the boundaries of the mixed phase
region in the $(T,\rho_B)$ plane. 
Since in general the charge chemical potential is related to the symmetry 
term of the EoS, \cite{baranPR},
$
\mu_3 = 2 E_{sym}(\rho_B) \frac{\rho_3}{\rho_B},
$ 
we expect critical and transition densities
rather sensitive to the isovector channel in the two phases.

In the hadron sector we use the NL-RMF
models, Sects.6-7, with
different structure of the isovector part:
i) $NL$, where no 
isovector
meson is included and the symmetry term is only given by the kinetic 
Fermi contribution, ii) $NL\rho$ when the interaction contribution of
an isovector-vector meson is considered and finally iii) $NL\rho\delta$
where also the contribution of an isovector-scalar meson is accounted for.
At high baryon densities the symmetry energy is increasing in a different
way with the various
choices, see Sect.6, and we will look at the effect on 
the phase transition.  

As already mentioned, in the quark phase we use the MIT-Bag Model, where 
the symmetry term is only given by the Fermi contribution. The Bag parameter
B is fixed for each baryon density to a constant, rather standard, value
$B^{1/4}=160MeV$, corresponding to a Bag Pressure of $85.7~MeV~fm^{-3}$.

\begin{figure}
\centering
\includegraphics[scale=0.27]{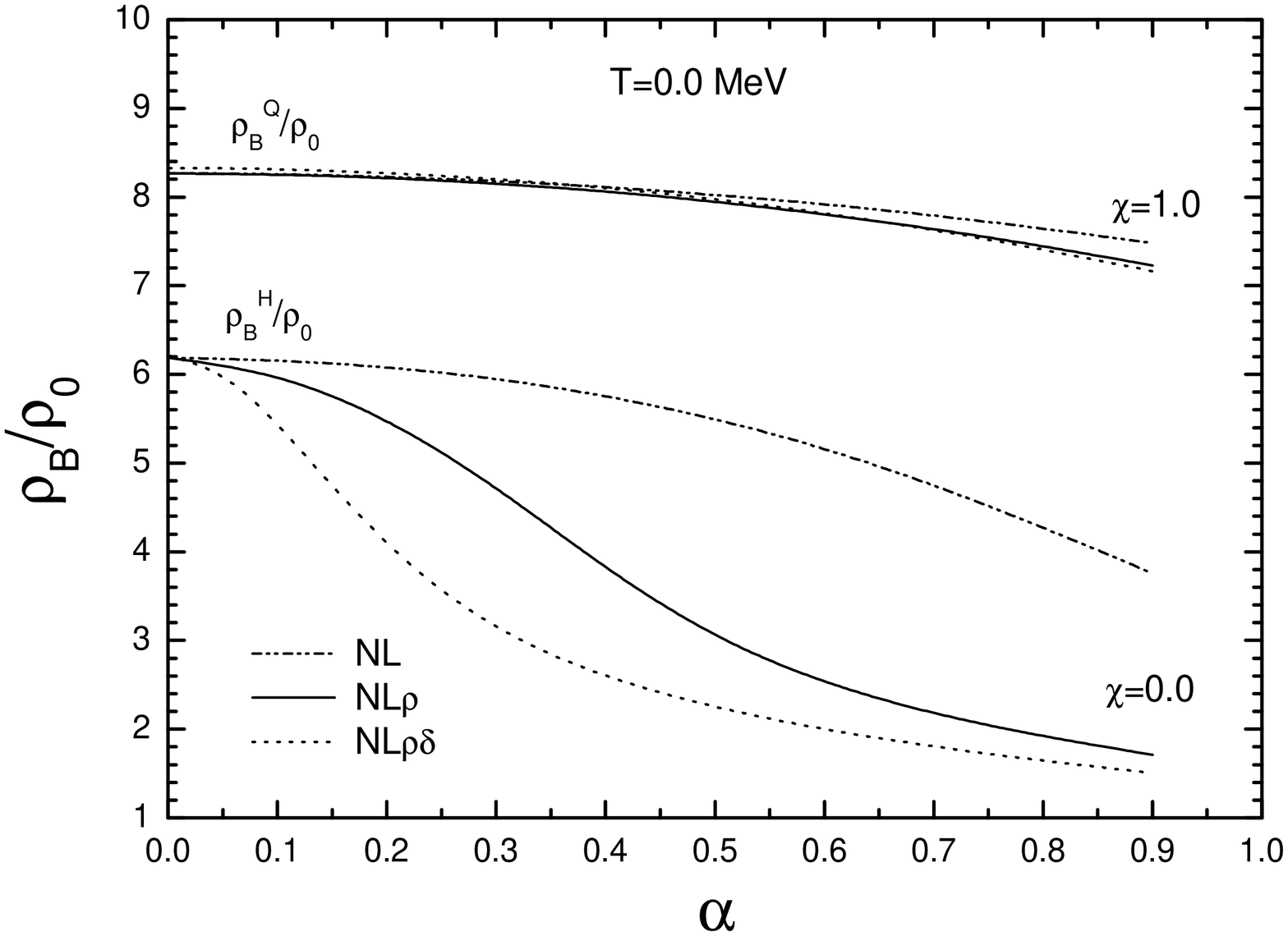}
\hskip -0.5cm
\includegraphics[scale=0.27]{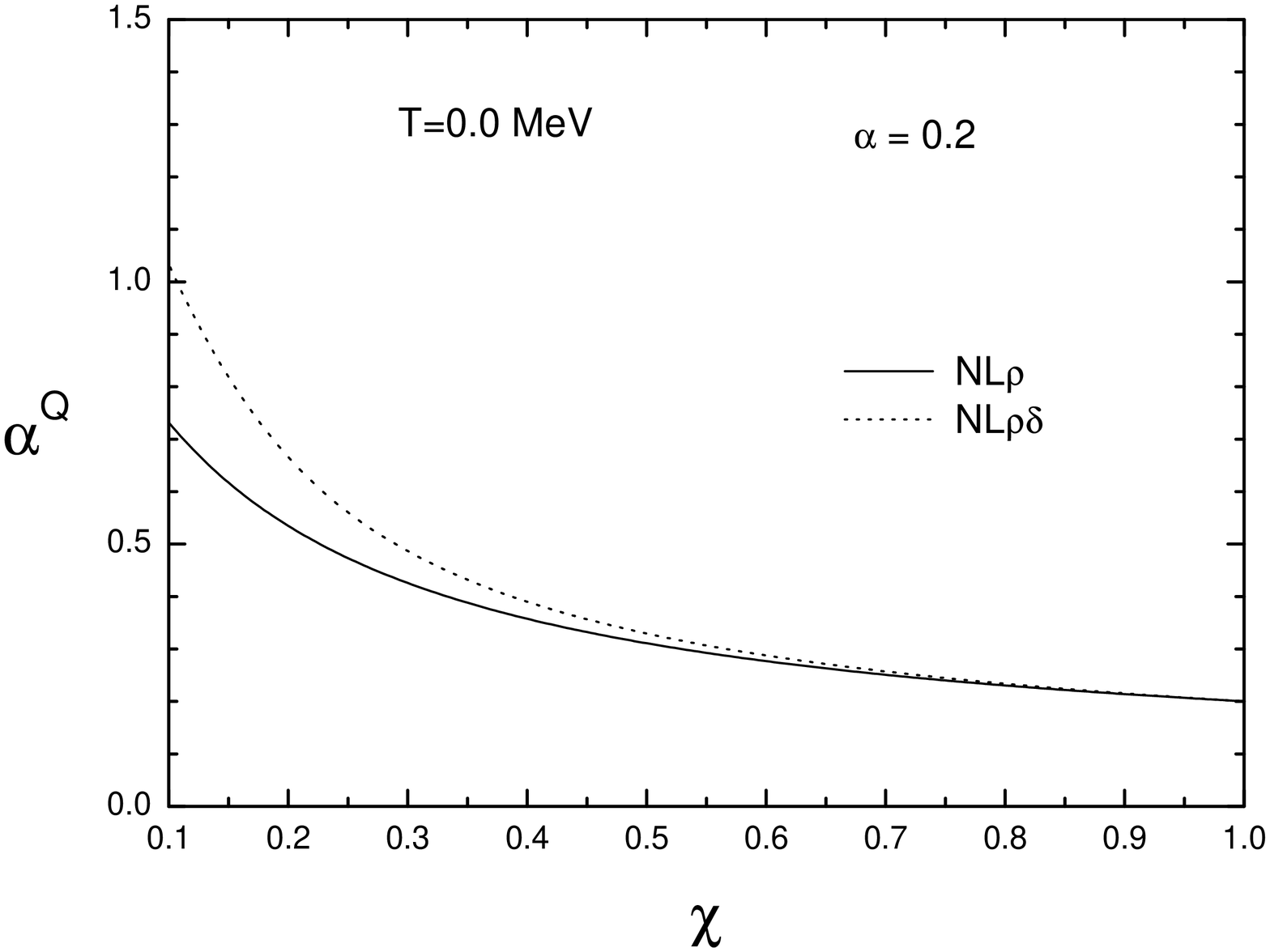}
\caption{Left Panel.
Dependence on the Hadron Symmetry Energy of the Lower ($\chi=0.0$) and
Upper ($\chi=1.0$) Boundaries of the Mixed Phase, at zero temperature,
vs. the asymmetry parameter.\\
Right Panel. Quark asymmetry in the mixed phase vs. the quark concentration
for asymmetric matter with $T=0$ and $\alpha=0.2$.
$NL\rho$ and $NL\rho\delta$ Effective Hadron Interactions are considered.
\\
Quark $EoS$: $MIT$ bag model with
$B^{1/4}$=160 $MeV$.
}
\label{boundaries}
\end{figure}

In general for each effective interactive Lagrangian we can simulate the 
solution of the highly non-linear system of Eqs.(\ref{gibbs}), via an
iterative minimization procedure, in order to determine the binodal 
boundaries and eventually
the Critical End Point (CEP)
($T_c,~\rho^B_c$) of the mixed phase.

A relatively simple calculation can be performed at zero temperature. The 
isospin effect (asymmetry dependence) on the Lower $(\chi=0.0)$ and 
Upper $(\chi=1.0)$ transition densities of the Mixed Phase are shown
in Fig.\ref{boundaries}(Left Panel) for various choices of the Hadron EoS. 
The effect
of a larger repulsion of the symmetry energy in the hadron sector, from
$NL$ to $NL\rho$ and to $NL\rho\delta$,
 is
clearly evident on the lower boundary with a sharp decrease of the transition 
density even at relatively low asymmetries.

Typical results for isospin effects on the whole binodal ``surface'' are 
presented 
in Fig.\ref{NLmix} for $\alpha=0.2$ asymmetric matter.
As expected, the lower boundary
of the mixed phase is mostly affected by isospin effects. In spite of the 
relatively small
total asymmetry, we clearly see a shift to the left of the
first transition boundary, in particular at low temperature, and a relatively 
``early'' Critical End Point. In fact
from the present results we cannot exclude a $CEP$ at higher temperature
and smaller baryon density, as suggested by other similar calculations
\cite{muller}. However we have also to mention some numerical problems.
The solution of the Gibbs conditions, the highly non-linear system
of Eqs.(\ref{gibbs}), is found through an iterative multi-parameter 
minimization procedure (the Newton-Raphson method). When we are close to the 
Critical End Point, for each $\chi$-concentration the baryon and isospin 
densities of the hadron and quark phases become very similar and we 
start to have problems in finding a definite minimum. 

In the following we will concentrate on properties of the mixed phase
mostly located at high density and relatively low temperature, well
described in the calculation and
  within the reach of
heavy ion collisions in the few AGeV range, see the discussion
around Fig.\ref{figUU}. 
We note that high intensity $^{238}U$ beams in this energy range
will be available in the first stage of the FAIR facility
\cite{fair,horst} and also at JINR-Dubna in the Nuclotron
first step of the NICA project \cite{sissakian08}.

\subsection{Inside the Mixed Phase of Asymmetric Matter}

\begin{figure}
\centering
\includegraphics[scale=0.27]{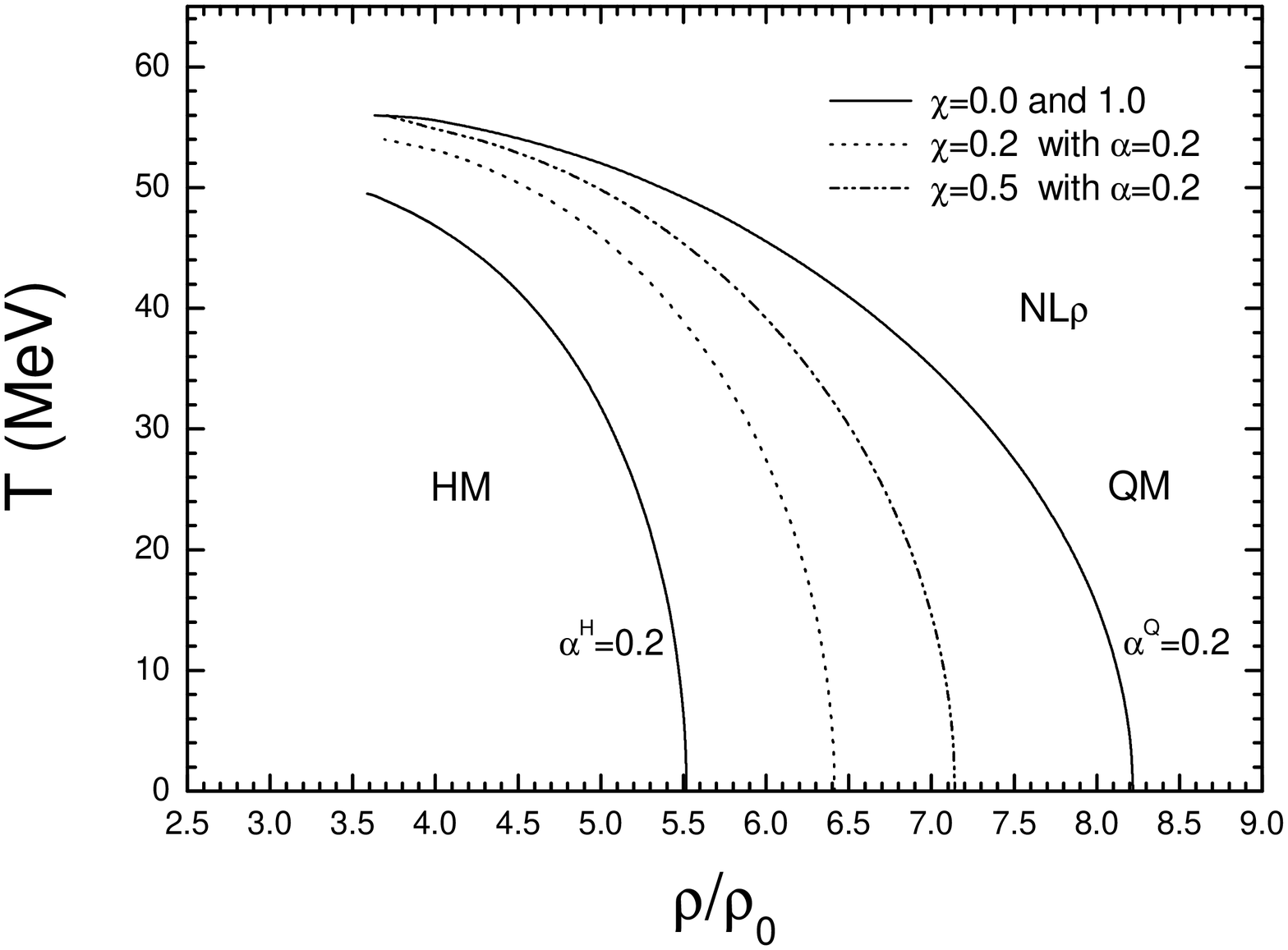}
\includegraphics[scale=0.27]{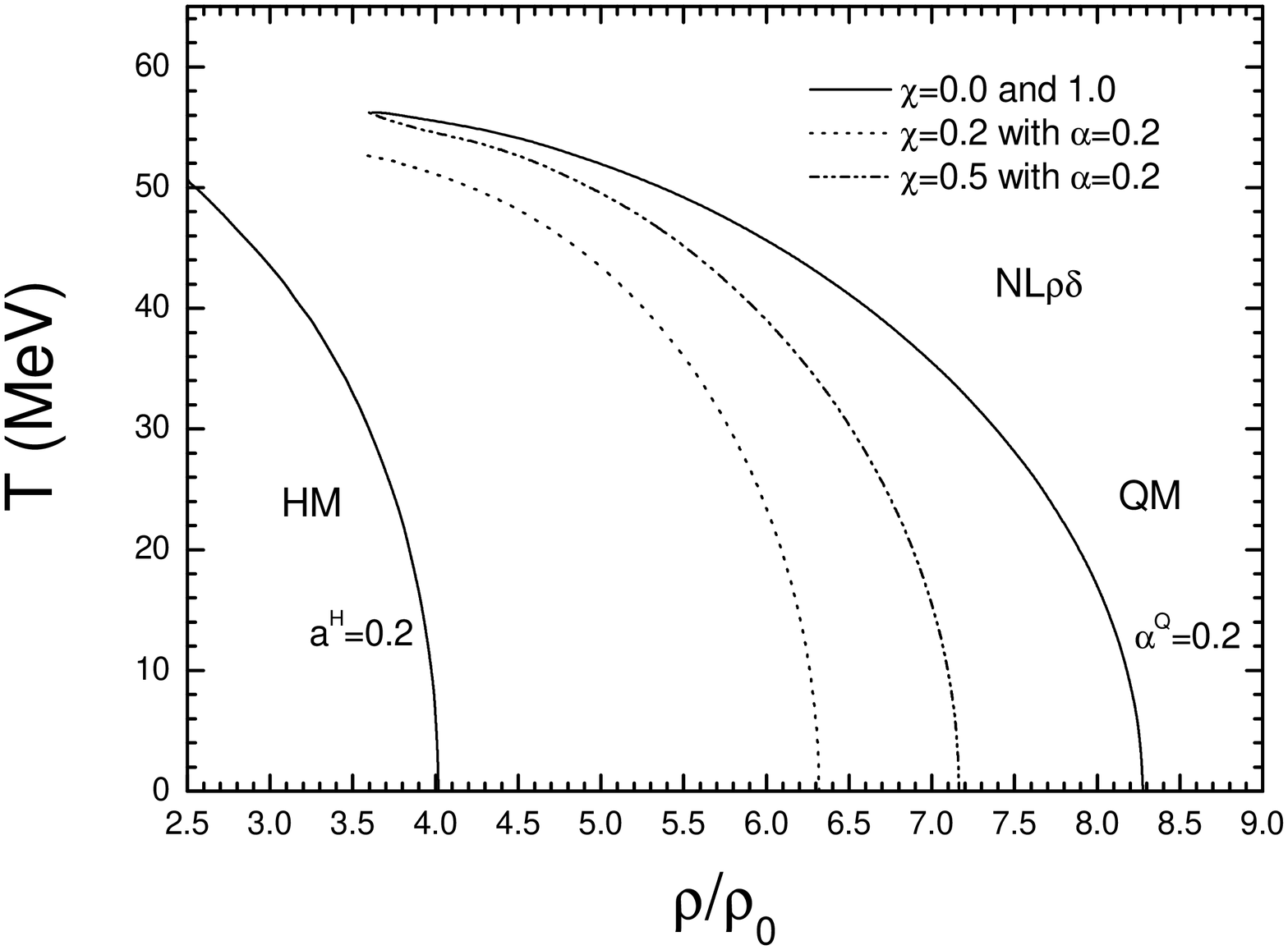}
\caption{Asymmetric $\alpha=0.2$ matter.
Binodal surface and ($T,\rho_B$)
curves for various quark concentrations
($\chi=0.2,0.5$) in the mixed phase.
Hadron $EoS$: $NL\rho$ Effective Interaction (Left Panel);
 $NL\rho\delta$ Effective Interaction (Right Panel).
Quark $EoS$: $MIT$ Bag model with
$B^{1/4}$=160 $MeV$.
}
\label{NLmix}
\end{figure}

For $\alpha=0.2$ asymmetric matter,
 in the Fig.\ref{NLmix}  we show also the ($T,\rho_B$)
curves inside the Mixed Phase corresponding to a $20\%$ and $50\%$ 
presence of the quark component ($\chi=0.2,0.5$), evaluated respectively
with the
two choices, $NL\rho$ (Left Panel) and $NL\rho\delta$ (Right Panel), of 
the symmetry interaction
in the hadron sector. We note, as also expected from Fig.\ref{boundaries},
that in the more repulsive $NL\rho\delta$ case the lower boundary is
much shifted to the left. However this effect is not so evident for the curve
corresponding to a $20\%$ quark concentration, and almost absent for the
$50\%$ case. The conclusion seems to be that for a stiffer symmetry term
in a heavy-ion collision at intermediate energies during the compression stage
we can have more chance
to probe the mixed phase, although in a region with small weight of the
quark component. 

Can we expect some signatures related to the subsequent
hadronization in the following expansion?

An interesting possibility is coming from the study of the asymmetry
$\alpha^Q$ in the quark phase. In fact since the symmetry energy is rather 
different in the two phases we can expect an Isospin Distillation 
(or Fractionation), very similar to the one observed in the Liquid-Gas 
transition in dilute nuclear matter \cite{baran98,baranPR,chomazPR}, this time 
with the larger isospin 
content in the higher density quark phase.

In Fig.\ref{boundaries} (Right Panel) we show the asymmetry $\alpha^Q$ in 
the quark phase
as a function of the quark concentration $\chi$ for the case with global
asymmetry $\alpha=0.2$ (zero temperature). The calculation is performed with 
the two choices of the symmetry term in the hadron sector. We see an impressive
increase of the quark asymmetry when we approach the lower boundary of the
mixed phase.
Of course the quark asymmetry recovers the global value 0.2 at the upper
boundary $\chi=1$. A simple algebraic calculation allows to evaluate the 
corresponding asymmetries of the hadron phase.
In fact from the charge conservation we have that for any $\chi$-mixture
 the global
asymmetry $\alpha$ is given by:

\begin{equation}\label{alphamix}
\alpha \equiv -\frac{\rho_3}{\rho_B} = 
\frac{(1-\chi)\alpha^H}{(1-\chi)+\chi \frac{\rho_B^Q}{\rho_B^H}} +
\frac{\chi \alpha^Q}{(1-\chi) \frac{\rho_B^H}{\rho_B^Q}+ \chi}
\end{equation}

For any $\chi$, from the calculated $\alpha^Q$ of Fig.\ref{boundaries} (Right)
and the $\rho_B^H,\rho_B^Q$ obtained from the Gibbs Eqs.(\ref{gibbs}),
we can get the correspondent asymmetry of the hadron phase $\alpha^H$.
 For a $20\%$ quark concentration
we have an $\alpha^Q/\alpha^H$ ratio around 5 for $NL\rho$ and around 20
for $NL\rho\delta$, more repulsive in the isovector channel
\cite{ditorojpg}. 
It is also interesting to compare the isospin content $N/Z$ of the high
density region expected from transport simulations without the Hadron-Quark
transition and the effective $N/Z$ of the quark phase in a 
$20\%$ concentration.
In the case of $Au+Au$ (initial $N/Z=1.5$) central collisions at $1AGeV$
in pure hadronic simulations we get in the high density phase a reduced
$N/Z \sim 1.2-1.25$ (respectively with $NL\rho\delta-NL\rho$ interactions)
due to the fast neutron emission \cite{ferini06,theo04}. In presence of the 
transition, the corresponding
isospin content of the quark phases is much larger, $N/Z=3$ for $NL\rho$ 
and $N/Z=5.7$ for $NL\rho\delta$. This is the $neutron~trapping$ effect.
We expect
a signal of such large asymmetries, coupled to a larger baryon density in 
the quark phase, in the subsequent 
hadronization. We could predict
an enhancement of the production of isospin-rich nucleon resonances and 
subsequent decays, i.e. an increase of $\pi^-/\pi^+$, $K^0/K^+$ yield ratios 
for reaction
products coming from high density regions, that could be selected looking
at large transverse
momenta, corresponding to a large radial flow.

 If such kinetic selection of particles from the mixed phase can 
really be successful also other mixed phase signatures would become available.
One is related to the general softening of the matter, due to the contribution
of more degrees of freedom, that should show up in the damping of collective 
flows \cite{csernai99}.

The azimuthal distributions (elliptic flows) will be
particularly affected since particles mostly retain their high transverse
momenta escaping along directions orthogonal to the reaction plane without
suffering much rescattering processes. 
A further signature could be the observation, for the selected particles, 
of 
the onset of a quark-number scaling of the elliptic flow: a property of 
hadronization
by quark coalescence that has been predicted and observed at RHIC energies,
i.e. for the transition at $\mu_B=0$ \cite{fries08}.

We finally remark that at higher temperature and smaller baryon chemical 
potential (ultrarelativistic collisions) the isospin effects discussed here 
are expected to vanish \cite{sissakian08}, even if other physics can enter 
the game and charge asymmetry effects are predicted also at $\mu_B=0$ and
$T \simeq 170~MeV$ \cite{kogut04,toublan05}.

\subsection{Isospin in Effective Quark Models}

From the above discussion it clearly appears that 
 the lack of explicit isovector
interactions in the quark sector could strongly affect the location 
of the phase transition in asymmetric matter and the related 
expected observables.
So it seems
extremely important to include the 
Isospin degree of freedom in any effective QCD dynamics. 
A first approach can be supplied by a two-flavor Nambu-Jona Lasinio ($NJL$) 
model \cite{NJL} 
where the isospin asymmetry can be included in a flavor-mixing picture 
\cite{frank03,buballa05}. These isospin effects are induced by a
determinant interaction related to the breaking of the axial symmetry.
The new  Gap Equations are like 
$M_i=m_i-4G_1 \Phi_i-4G_2 \Phi_j$, $i \not= j,(u,d)$ , where the 
$\Phi_{u,d}=<\bar u u>,<\bar d d>$ are the two (negative) condensates 
and $m_{u,d}=m$
the (equal) current masses. Introducing explicitily a flavor mixing, i.e.
the dependence of the constituent mass of a given flavor to both condensate,
via $G_1=(1-\eta) G_0, G_2= \eta G_0$ we have the coupled equations
\begin{eqnarray}
&&M_u=m - 4 G_0 \Phi_u + 4 \eta G_0 (\Phi_u - \Phi_d), \nonumber \\ 
&&M_d=m - 4G_0 \Phi_u + 4 (1-\eta) G_0 (\Phi_u - \Phi_d).
\label{mix}
\end{eqnarray}
For $\eta=1/2$ we have back the usual NJL ($M_u=M_d$), while small/large
mixing is for $\eta \Rightarrow 0$/$\eta \Rightarrow 1$ respectively.

In neutron rich matter $\mid \Phi_d \mid$ decreases more rapidly due to the 
larger $\rho_d$ and so $(\Phi_u -\Phi_d)<0$. In the ``realistic'' small mixing 
case, see also \cite{frank03,shao06}, we will get a definite $M_u>M_d$ 
splitting at high baryon density (before the chiral restoration). 
This expectation is nicely confirmed by a full calculation \cite{plum_tesi}
of the coupled gap equations with standard parameters (same as in 
ref.\cite{frank03}). 

All that can indicate a more fundamental confirmation of the $m^*_p>m^*_n$ 
splitting in the hadron phase,
as suggested by the effective $QHD$ model with the isovector scalar $\delta$ 
coupling, see \cite{baranPR,liubo02} and Sects.5, 6.

However such isospin mixing effect results in a very small variation
of the symmetry energy in the quark phase, still related only to the Fermi
kinetic contribution. 
 In fact this represents just a very first step towards a more complete
treatment of isovector interactions in effective quark models, of large 
interest for the discussion of the phase transition at high densities.
We remind that confinement 
is still 
missing in these mean field approaches. 

More generally starting from the QCD Lagrangian one can arrive to an 
effective color current-current interaction where an expansion in various 
components can provide isovector contributions \cite{weise07}.
We also notice the evidence that chiral symmetry restoration
is favored in systems with large neutron excess \cite{kaiser09}. 

A very interesting point has been recently suggested about the possibility
of a quark matter formed in a color superconducting state \cite{pagliara10}.
A strong symmetry repulsion comes from the fact that equal densities of up 
and down quarks are energetic favored to allow the formation of a 
diquark condensate. This will be in competition with a decreasing of the
diquark pairing at large isospin asymmetries. 
Exciting new perspectives are open up.

In conclusion the aim of this project is twofold:
\begin{itemize}
\item
{To stimulate new experiments on isospin effects in heavy ion collisions 
at intermediate energies (in a few $AGeV$ range) with  attention
to the isospin content of produced particles and to elliptic flow properties,
in particular for high-$p_t$ selections.}
\item{To stimulate more refined models of effective Lagrangians for
non-perturbative QCD, where isovector channels are consistently accounted for.}
\end{itemize}

\section{Perspectives and Suggested Observables}

We have shown how dissipative Heavy Ion Collisions with charge asymmetric 
isotopes provide a valuable, in fact unique, tool to extend our knowledge 
of the Nuclear Matter Phase Diagram along the "third" isovector component,
$\rho_3=\rho_n-\rho_p$. Due to the low asymmetries reached in terrestrial 
laboratories, even with unstable beams, it is difficult to disentangle 
density and momentum dependence of the in-medium symmetry potentials from 
the corresponding properties of the isoscalar interactions. Therefore our 
discussion has been mainly focused on exclusive experiments, where suitable 
correlations can better isolate isovector effects.

The fundamental microscopic tool to relate isospin properties of the reaction
products to the underlying interaction is the transport theory to describe the 
evolution of the reaction dynamics. We must remark that isoscalar global
properties of the collision dynamics (like stopping power and density 
variations, pre-equilibrium emissions and main reaction mechanisms) should
be correctly reproduced otherwise also the isovector predictions will be not 
fully reliable. We remind that one of the most important information is about
the density dependence of the symmetry energy away from saturation. This is 
in fact a basic requirement for a comparison of different transport models.

We have used a non-relativistic and 
relativistic Stochastic Mean Field approach. In this way we can account for 
the very important Mean Field Dynamics coupled to Fluctuation Terms that can 
lead to fragment formation in mechanical/chemical unstable regions, and in 
general to the onset of various other instabilities, as well as to the widths 
in the distributions of the measured quantities.

We have selected some relevant features of the Isospin Dynamics:
\begin{itemize}
\item{
a) Collective Isospin Equilibration (Sect.3)}
\item{
b) Isospin Diffusion due to asymmetry gradients (Sect.4.1)}
\item{
c) Isospin Distillation in multifragmentation (Sect.4.2)}
\item{
d) Isospin Migration due to density gradients (Sect.4.3)}
\item{
e) Isospin flows and isospin effects on nucleon, cluster and meson
   production at high energies (Sects. 5 and 7)}
\item{
f) Isospin effects on the Mixed Phase in the hadron-quark transition
   at high baryon density (Sect.8)}
\end{itemize}

In correspondence we have suggested the more promising exclusive
measurements:

\noindent
1. Angular distribution of the prompt dipole radiation, measured with
   high-intensity n-rich ions, like $^{132}Sn$.

\noindent
2. Correlation between isospin diffusion and total energy loss, which gives 
   a direct measure of the interacting time.

\noindent
3. Correlation between fragment isospin content and radial velocity, possibly 
   with a reconstruction of the primary fragments (in particular the neutron 
   emission).

\noindent
4. Angular, mass and velocity correlations of the isospin properties of 
   the neck fragments.

\noindent
5. Transverse momentum analysis of the isospin content of nucleon, clusters 
   and mesons emitted at mid rapidity in intermediate energy collisions, and 
   of the corresponding collective flows.

\noindent
6. High $p_t$ study of the quenching of hadron elliptic flows joined to some 
   evidence of the "neutron trapping" effect (enhancement of production of 
   isospin-rich nucleon resonances and subsequent decays) as indicator of the 
   formation of a mixed hadron-quark phase.

From the present available data at the Fermi energies we can deduce a rather 
stiff (close to a linear behavior) symmetry energy below/around the 
saturation density, from points b), c), d). Similar indications are appearing
from point e) concerning the higher density regions. Here the isospin effects
on the momentum dependence (splitting of the nucleon effective masses) seem
to be also relevant.
Very nice perspectives are opening with the new facilities, even for unstable 
beams, soon operating at higher energies.

\subsection*{Acknowledgments}

On some of the topics presented here we warmly thank the fruitful and 
pleasant collaboration of very nice people: G.Ferini, Th.Gaitanos, V.Giordano,
B.Liu, F.Matera, S.Plumari, V.Prassa, C.Rizzo, J.Rizzo, M.Zielinska-Pfabe
and H.H.Wolter. We acknowledge several inspiring discussions with theory and 
experiment colleagues: B.Borderie, Ph.Chomaz, P.Danielewicz, E.De Filippo, 
E.Galichet,
C.M.Ko, B.A.Li, Qinfeng Li, W.G.Lynch, A.Pagano, J.Randrup, W.Reisdorf, 
M.F.Rivet, 
P.Russotto, D.V.Shetty, C.Simenel, H.Stoecker, W.Trautmann, M.B.Tsang, 
J.Wilczynski
 and S.J.Yennello.

 One of the authors, V. B. thanks for warm hospitality at Laboratori
Nazionali del Sud, INFN. This work was supported in part by the Romanian
Ministery for Education and Research under the CNCSIS contract 
PN II ID-946/2007
and CNMP grant PNII-Partnerships No. 71-073/2007-PROPETHAD.

\vspace{0.5cm}



\begin{thebibliography}{99}
\bibitem{imbalance}
We note that the Imbalance Ratio appears also not dependent on pre-equilibrium
emissions as well as on secondary decays \cite{tsang04,rizzo08}.
\bibitem{nonlinear}
We use a ``Minimal'' Non-Linear (NL) Lagrangian, with self-interacting
contributions inserted only for the isoscalar/scalar $\sigma$ field
in order to reproduce a relatively $soft$ EoS for symmetric matter, in
agreement with monopole and flow data.
\bibitem{couplings}
The meson-nucleon couplings are considered not density dependent and
chosen in order to reproduce good saturation properties and average
Dirac-Brueckner-Hartree-Fock predictions at high density 
\cite{hofmann01,goegelein08}, see details in refs.\cite{liubo02,baranPR}.
\bibitem{nlrho}
We note that our $NL\rho$ effective 
Lagrangian is
very close to other widely used relativistic effective models, e.g. see
the $GM3$ of ref.\cite{GlendenningPRL18}
and the $NL3$ interaction of P.Ring
and collaborators \cite{vretenar03}, which has also given good nuclear 
structure 
results, even for exotic nuclei.
\end{thebibliography}
\end{document}